# Thermodynamic extension of density-functional theory.
# II. Finite-temperature ensemble spin-density functional theory

Robert Balawender[1,2,a] and Andrzej Holas[1]

1) Institute of Physical Chemistry of Polish Academy of Sciences, Kasprzaka 44/52, PL-01-224 Warsaw, Poland
2) Eenheid Algemene Chemie (ALGC), Vrije Universiteit Brussel, Pleinlaan 2, B-1050 Brussels, Belgium

Abstract.

The formalism developed in the first paper of the series [Ref.[1]] is applied to the thermodynamic spin-density functional theory (TSDFT) by considering two thermodynamic systems characterized by: (i) three global observables (the energy, the total electron number and the spin number), (ii) one global observable (the internal electron energy) and two local (position-dependent) observables (the total electron density and the spin density). The two-component potential of the many-electron system of interest consists a scalar external potential and a collinear magnetic field (coupled only with the spin operator). Various equilibrium characteristics – the state functions – of two systems are defined and investigated. Conditions for the equivalence between two systems (the same equilibrium density matrix required) are derived and thoroughly discussed. The applicability of the Hohenberg-Kohn theorem is extended to the TSDFT. Obtained results provide a rigorous mathematical foundation for future derivation of the zero-temperature limit of this theory and determination of its chemical reactivity indices.

---

[a] Author to whom correspondence should be addressed,
rbalawender@ichf.edu.pl

# I. INTRODUCTION

Conceptual density-functional theory (DFT) offers a perspective on the interpretation and prediction of experimental and theoretical reactivity data based on a series of response functions to perturbations in the number of electrons and/or external potential.[2-10] This approach has enabled a sharp definition and then computation, from first principles, of a series of well-known but sometimes vaguely defined chemical concepts such as electronegativity, hardness, and Fukui function. Identification of chemical concepts and principles in terms of derivatives with respect to the electron density (local chemical reactivity descriptors) or with respect to the total electron number (global properties)[4-7,11-13] requires an extension of the density domain to ensemble density which integrates to a positive *real* ("fractional") electron number. Despite a widespread popularity and success of DFT, its applications can still suffer from large pervasive errors that cause qualitative failures in predicted properties.[14-17] It is due to violations by currently used approximate functionals of the exact conditions that should by satisfied by DFT.[18-21] Only functionals defined in the thermodynamic approach satisfy these conditions: they are continuous, convex and size consistent.[18,20,22] The formal foundation of thermodynamic density functional theory was given long ago by Mermin.[23] A thermodynamic formulation based on a general grand canonical ensemble was given by Rajagopal.[24] Other generalizations have been reported[25-36] too. DFT was extended to include spin-dependent potentials by von Barth and Hedin[37] and independently by Pant and Rajagopal.[38] The case of temperature dependence in the spin density-functional theory (SDFT) was considered by Gunnarsson and Lunqvist.[39] Formulation of SDFT based on the constrained-search approach[8,40] was proposed.[41-43] The existence of the functional derivative in thermodynamic extension of the SDFT was proved.[44]

In the first paper[1] of the series, hereafter referred to as Paper I, the formulation of the equilibrium state of a many-electron system in terms of an ensemble (mixed-state) density matrix operator, which applies the maximum entropy principle combined with the use of the basic

Massieu function, was presented. Based on the effective action formalism, the convexity and concavity properties of the basic Massieu function, its Legendre transforms (Massieu functions) and the corresponding Massieu-Planck transforms (Gibbs-Helmholtz functions) for various ensembles were determined. Since the arguments used in Paper I were general (the maximum entropy principle, the effective action formalism), obtained results provide a rigorous mathematical foundation for generalizing the SDFT to finite temperature and to fractional number of electrons and spins in the framework of the ensemble approach, and also for analyzing the zero-temperature limit of this extension. The general quantum statistical mechanics approach of Paper I will be confined here to a thermodynamic SDFT (TSDFT) by adopting the constrained search approach of Levy.[40] This approach has been applied previously to pure states, to an ensemble of pure states with fixed integral electron number,[8,45] an to an ensemble with a fluctuating electron number.[46,47] In Sec.II we introduce the basic spin formalism and also two thermodynamic systems to study their equilibrium state. The first of them is a system of three global observables (in which the average energy, the average total electron number and the average spin number can be determined), the second is a system of one global observable (the average internal electron energy can be determined) and two local (position-dependent) observables (the average total density and the average spin density determined). In Sec.III and Sec.IV we discuss equilibrium characteristics of the first and second system, respectively, for various ensembles. As the formal structure of the phenomenological thermodynamics can be formulated either in the energy representation or in the equivalent dimensionless representation,[48,49] – the entropy representation (i.e., in terms of the $(M,m)$ Gibbs-Helmholtz functions or the corresponding $(M,m)$ Massieu functions, respectively, defined in Paper I), the equilibrium state properties are discussed in both representations. In Sec.V we impose the requirement for two systems of interest to have the same equilibrium density matrix operator, and then perform an analysis of this equivalence between two systems for the spin-grand-canonical ensemble and the spin-canonical ensemble. Furthermore, the relations obtained for the two systems are used in Sec.VI to formulate an extension of the Hohenberg-Kohn theorem to TSDFT expressed in two equivalent representations. Finally, in Sec.VII we present our conclusions.

*Note*: Atomic units are used throughout the paper. As in Paper I, a column matrix — a vector in *M*-dimensional space — is denoted by a variable with a subscript which indicates an ordered set of indices, e.g. $x_\mathsf{F}$. A row matrix is the transpose of a column matrix, $x_\mathsf{F}^\mathrm{T} = (x_1 x_2 ... x_M)$. Here $\mathsf{F}$ means the full range of indices in the set, $\mathsf{F} = \{1,2,...,M\}$. It may be partitioned into subsets, e.g. the lower range $\mathsf{L} = \{1,2,...,m\}$ and the upper one $\mathsf{U} = \{m+1,...,M\}$, $\mathsf{F} = \mathsf{L} \cup \mathsf{U}$. The corresponding block structure of a row matrix is denoted as $x_\mathsf{F}^\mathrm{T} = \left( x_\mathsf{L}^\mathrm{T} \mid x_\mathsf{U}^\mathrm{T} \right)$. For typographic reasons, the corresponding column $x_\mathsf{F} = \left( x_\mathsf{L}^\mathrm{T} \mid x_\mathsf{U}^\mathrm{T} \right)^\mathrm{T}$ may be alternatively denoted as $x_\mathsf{F} = (x_\mathsf{L}, x_\mathsf{U})$. A square or rectangular matrix is distinguished by two set subscripts, e.g. a square matrix $\Phi_\mathsf{FF}$ can be composed of its blocks: square ones $\Phi_\mathsf{LL}, \Phi_\mathsf{UU}$, and rectangular ones $\Phi_\mathsf{LU}$ and $\Phi_\mathsf{UL}$, and so on for higher-rank matrices like $\Psi_\mathsf{FFF}$ etc. The notation for specific vectors is also introduced (see Section II): a two-component column vector (2vec) of quantities like the density, the external potential, electron number, sources, etc., is denoted by a bold italic letter: $\boldsymbol{\rho}$, $\boldsymbol{v}$, $\boldsymbol{\mathcal{N}}$, $\boldsymbol{\mu}$, etc.; the "transpose" mark at 2vec is omitted, e.g., the scalar product denoted $\boldsymbol{\mu}\boldsymbol{\mathcal{N}}$ means $\boldsymbol{\mu}^\mathrm{T}\boldsymbol{\mathcal{N}}$ or $\boldsymbol{\mathcal{N}}^\mathrm{T}\boldsymbol{\mu}$; a vector in the configuration space is denoted by a non-italic, bold letter, e.g. $\mathbf{r}$, $\mathbf{R}_a$. The integral functional $\int[...]$ is introduced to represent the integration over the whole space of a function of a spatial variable, $\int[f] \equiv \int d\mathbf{r}\, f(\mathbf{r})$. The abbreviation "w.r.t." means "with respect to", "fn." means "function", "fnl." means "functional".

## II. OPERATORS AND SYSTEMS SUITABLE FOR SPIN-DENSITY FUNCTIONALS

We consider a molecule as a system of many electrons moving in a scalar external potential field $v_{ext}(\mathbf{r})$ (usually due to the clamped nuclei: $v_{ext}(\mathbf{r}) = -\sum_a Z_a/|\mathbf{r} - \mathbf{R}_a|$), and in a collinear magnetic field (here multiplied by the Bohr magneton) $(B_x, B_y, B_z) = (0, 0, B_z(\mathbf{r}))$. The system is described by the non-relativistic, time-independent Hamiltonian

$$\hat{H}[v] = \hat{T} + \hat{U} + \hat{V}[v] = \hat{F}_{int} + \hat{V}[v] \tag{1}$$

with $\hat{T}, \hat{V}$ — the one-body operators of the kinetic energy and the interaction energy with external fields, respectively, and $\hat{U}$ — the two-body operator of the electron-electron interaction energy. The Hamiltonian can be partitioned alternatively into the *internal* energy operator $\hat{F}_{int} = \hat{T} + \hat{U}$ and the *external* energy operator $\hat{V}[v]$. All operators written in the second-quantized formulation act in the Fock space. The scalar external potential needs to be sufficiently attractive to allow $\hat{H}$ to have bound states for $\mathcal{N}$-electron systems, $1 \leq \mathcal{N} \leq \mathcal{N}_0[v_{ext}]$ (the maximum number for this potential). The magnetic field is assumed to be nonuniform, $B_z(\mathbf{r}) \neq \text{const}$, to avoid noninvertibility of the mapping between potentials and densities at the zero temperature limit.[30,50-56] The role of the magnetic field is restricted only to the interaction term with the electron spin, while the vector potential of this field in the kinetic energy operator $\hat{T}$ and the dipolar interaction between spins are neglected. Such Hamiltonian is appropriate for describing approximately a molecule in a weak magnetic filed or its spontaneous magnetization occurring in the absence of the magnetic field (i.e., treated as the limit $B_z(\mathbf{r}) \to 0$).

The interaction operator can be written as $\hat{V}[v] = \int [v \hat{\rho}]$, i.e., as an integral of the scalar product of 2vecs: the density operator $\hat{\rho}(\mathbf{r}) = (\hat{\rho}_\uparrow(\mathbf{r}), \hat{\rho}_\downarrow(\mathbf{r}))$ [defined in terms of the field operators as $\hat{\rho}_\sigma(\mathbf{r}) = \hat{\psi}^\dagger(\mathbf{r},\sigma)\hat{\psi}(\mathbf{r},\sigma)$] and the potential $\mathbf{v}(\mathbf{r}) = (v_\uparrow(\mathbf{r}), v_\downarrow(\mathbf{r})) \equiv (v_{ext}(\mathbf{r}) + B_z(\mathbf{r}), v_{ext}(\mathbf{r}) - B_z(\mathbf{r}))$. The electron-number operator 2vec is $\hat{\mathcal{N}} = \int[\hat{\rho}] = (\hat{\mathcal{N}}_\uparrow, \hat{\mathcal{N}}_\downarrow)$. The popular notation for $\{\alpha, \beta\}$ spin labels $\sigma \in \{\uparrow, \downarrow\}$ is not used to avoid confusion with a different application of the letters $\alpha$ and $\beta$. The expectation values of the operators will be denoted $\boldsymbol{\rho}(\mathbf{r}) = (\rho_\uparrow(\mathbf{r}), \rho_\downarrow(\mathbf{r}))$ and $\mathcal{N} = (\mathcal{N}_\uparrow, \mathcal{N}_\downarrow)$. It should be noted that by choosing the magnetic field to be collinear, $(B_x, B_y, B_z) = (0, 0, B_z(\mathbf{r}))$, we introduce the so called *diagonal* spin dependence to DFT, because only diagonal elements $\hat{\rho}_{\sigma\sigma}(\mathbf{r}) \equiv \hat{\rho}_\sigma(\mathbf{r})$ of the $2 \times 2$ spin-density matrix $\hat{\rho}_{\sigma\tau}(\mathbf{r}) = \hat{\psi}^\dagger(\mathbf{r},\sigma)\hat{\psi}(\mathbf{r},\tau)$ are involved.

Besides the above definitions, it proves convenient to introduce alternative forms of the potential and density 2vecs that leave the operator $\hat{V}[\mathbf{v}] = \int[\mathbf{v}\hat{\boldsymbol{\rho}}]$ unchanged, namely $\mathbf{v}(\mathbf{r}) = (v_{ext}(\mathbf{r}), B_z(\mathbf{r}))$, and $\hat{\boldsymbol{\rho}}(\mathbf{r}) = (\hat{\rho}_N(\mathbf{r}), \hat{\rho}_S(\mathbf{r})) \equiv (\hat{\rho}_\uparrow(\mathbf{r}) + \hat{\rho}_\downarrow(\mathbf{r}), \hat{\rho}_\uparrow(\mathbf{r}) - \hat{\rho}_\downarrow(\mathbf{r}))$ with the corresponding $\hat{\mathcal{N}} = (\hat{\mathcal{N}}, \hat{\mathcal{S}}) \equiv (\hat{\mathcal{N}}_\uparrow + \hat{\mathcal{N}}_\downarrow, \hat{\mathcal{N}}_\uparrow - \hat{\mathcal{N}}_\downarrow)$, where $\hat{\mathcal{N}}$ is the (total) particle-number operator, while $\hat{\mathcal{S}} = 2\hat{S}_z$ will be termed the spin-number operator [$\hat{S}_z$ is the z-component of the total-spin vector operator $(\hat{S}_x, \hat{S}_y, \hat{S}_z)$ of the many-electron system]. The corresponding expectation values are $\boldsymbol{\rho}(\mathbf{r}) = (\rho_N(\mathbf{r}), \rho_S(\mathbf{r}))$, $\mathcal{N} = (\mathcal{N}, \mathcal{S})$. The latter (alternative) forms of all 2vecs will be our principal ones for presenting the results. Separate components of any 2vec will labeled with subscripts N and S (except $(\mathcal{N}_N, \mathcal{S}_S)$ denoted as $(\mathcal{N}, \mathcal{S})$ and $(v_N(\mathbf{r}), v_S(\mathbf{r}))$ denoted as $(v_{ext}(\mathbf{r}), B_z(\mathbf{r}))$). Since the assumed $\hat{H}$ commutes with $\hat{\mathcal{N}}$, the pure eigenstate of this $\hat{H}$ can be labeled also by $\mathcal{N}$ – the admissible 2vec quantum number such that for any particle number $\mathcal{N} \in \{0, 1, 2, ...\}$ there are $(\mathcal{N} + 1)$ possible values of spin numbers $S \in \{-\mathcal{N}, -\mathcal{N}+2, \ldots, \mathcal{N}-2, \mathcal{N}\}$, which are all even (when $\mathcal{N}$ is even) or all odd (when $\mathcal{N}$ is odd). It should be noted that the TSDFT, discussed in the present paper, can be reduced to the more familiar spinless version by neglecting the second component of each 2vec used below. The formal justification of such reduction will be discussed in the paragraph after Eq.(19), and also in one after Eq.(55).

We will consider the equilibrium state of two thermodynamic systems, each characterized by its specific set of observables. The *first*, $\{\hat{H}, \hat{\mathcal{N}}\}$ system (of three global observables) satisfies

$$\text{Tr}\,\hat{\Gamma}_{\text{eq}}\hat{H} = E, \qquad \text{Tr}\,\hat{\Gamma}_{\text{eq}}\hat{\mathcal{N}} = \mathcal{N}, \qquad (2)$$

when it is fully isolated. Here $E$ and $\mathcal{N} = (N, S)$ are the average energy and the average electron-number 2vec, respectively. When fully open, the system is characterized by the conjugate external sources (Lagrange multiplies) $\beta, \alpha = (\alpha_N, \alpha_S)$. Note that this system depends implicitly on the potential 2vec $v(\mathbf{r})$ because $\hat{H} = \hat{H}[v]$.

The *second*, $\{\hat{F}_{\text{int}}, \hat{\rho}(\mathbf{r})\}$ system (of one global observable and two local ones) satisfies

$$\text{Tr}\,\hat{\Gamma}_{\text{eq}}\hat{F}_{\text{int}} = F_{\text{int}}, \qquad \text{Tr}\,\hat{\Gamma}_{\text{eq}}\hat{\rho}(\mathbf{r}) = \rho(\mathbf{r}), \qquad (3)$$

when it is fully isolated. Here $F_{\text{int}}$ and $\rho(\mathbf{r}) = (\rho_N(\mathbf{r}), \rho_S(\mathbf{r}))$ are the average internal electron energy and the average density 2vec. When fully open, the system is characterized by the conjugate external sources $b, w(\mathbf{r}) = (w_N(\mathbf{r}), w_S(\mathbf{r}))$. Fns./fnls. belonging to the first (second) system will be distinguished by the arc (tilde) over their symbols, e.g. $\breve{\Lambda}$ ($\tilde{\Lambda}$).

Starting from the basic Massieu fn., we can define many ensembles (characterized by the $(M, m)$ Massieu fns.) through variables reordering and Legendre transforms. For the $\{\hat{H}, \hat{\mathcal{N}}\}$ system, out of all possible 24 transforms, only 8 are independent (i.e. there is 8 distinct ensembles). They are presented in Fig. 1. As an example of equivalent transforms, for $\{\hat{H},\hat{\mathcal{S}},\hat{\mathcal{N}}\} \Rightarrow (j)$ we have $\breve{\Theta}^{3,1}_{(j)}[\beta,\mathcal{S},\mathcal{N}] = \breve{\Theta}^{3,1}[\beta,\mathcal{N},\mathcal{S}]$.

## III. EQUILIBRIUM CHARACTERISTICS OF $\{\hat{H},\hat{\mathcal{N}}\}$ SYSTEM

When fully open, the system is characterized by the conditions operator

$$\hat{\tilde{O}} \equiv \hat{\tilde{O}}[\beta,\boldsymbol{\alpha};\boldsymbol{v}] = -\beta\hat{H}[\boldsymbol{v}] - \boldsymbol{\alpha}\hat{\mathcal{N}}, \qquad (4)$$

where $\beta$ and $\boldsymbol{\alpha} = (\alpha_N,\alpha_S)$ are the Lagrange multipliers, the sources conjugate to $\hat{H}$ and $\hat{\mathcal{N}}$. We conveniently reduced the general dependence of $\hat{\tilde{O}}$ on $\{\hat{H},\hat{\mathcal{N}}\}$ to the dependence on the 2vec $\boldsymbol{v}(\mathbf{r})$ — the only system-specific characteristic of the observable operators $\{\hat{H},\hat{\mathcal{N}}\}$; their remaining constituents are universal Fock operators (the electron-electron interaction Coulombic potential of $\hat{F}_{\text{int}}$ is considered to be a universal function too). The source $\beta$ in Eq.(4) plays the role of the reciprocal temperature (in energy unit).

Due to $\int[\hat{\rho}(\mathbf{r})] = \hat{\mathcal{N}}$, it is easy to verify that $\hat{\tilde{O}}[\beta,\boldsymbol{\alpha}-\beta\mathbf{c};\boldsymbol{v}+\mathbf{c}] = \hat{\tilde{O}}[\beta,\boldsymbol{\alpha};\boldsymbol{v}]$ for a constant 2vec $\mathbf{c} = (c_N,c_S)$. This means that properties of a fully open $\{\hat{H},\hat{\mathcal{N}}\}$ system remain the same if the constant shift $\mathbf{c}$ of the 2vec external potential $\boldsymbol{v}(\mathbf{r})$ is accompanied by the shift $(-\beta\mathbf{c})$ of the 2vec source $\boldsymbol{\alpha}$. In other words, properties of the open systems depend solely on the combination $(\boldsymbol{v}(\mathbf{r}) + \beta^{-1}\boldsymbol{\alpha})$, although their externall potentials and sources (conjugate to $\hat{\mathcal{N}}$) may differ by constants.

According to Eq.(I.8) (i.e., Eq.(8) of Paper I), the equilibrium density matrix (eq-DM) operator is

$$\hat{\breve{\varGamma}}_{\mathrm{eq}}[\beta,\boldsymbol{\alpha};\boldsymbol{v}] = \exp\left(\hat{\breve{O}}^{[\beta,\boldsymbol{\alpha};\boldsymbol{v}]}\right) \Big/ \mathrm{Tr}\exp\left(\hat{\breve{O}}^{[\beta,\boldsymbol{\alpha};\boldsymbol{v}]}\right) \tag{5}$$

(when convenient, the bracketed arguments, like $[\beta,\boldsymbol{\alpha};\boldsymbol{v}]$ here, may be placed alternatively as a superscript). It serves for evaluation of the expectation values of operators, e.g.,

$$\breve{E}[\beta,\boldsymbol{\alpha};\boldsymbol{v}] = \mathrm{Tr}\,\hat{\breve{\varGamma}}_{\mathrm{eq}}^{[\beta,\boldsymbol{\alpha};\boldsymbol{v}]}\hat{H}[\boldsymbol{v}], \tag{6}$$

$$\breve{\boldsymbol{\mathcal{N}}}[\beta,\boldsymbol{\alpha};\boldsymbol{v}] = \mathrm{Tr}\,\hat{\breve{\varGamma}}_{\mathrm{eq}}^{[\beta,\boldsymbol{\alpha};\boldsymbol{v}]}\hat{\boldsymbol{\mathcal{N}}}. \tag{7}$$

The characteristic state function for this ensemble (see Eq.(I.10)) – the basic Massieu fn. –

$$K[\beta,\boldsymbol{\alpha};\boldsymbol{v}] \equiv \breve{\Theta}^{3,3}[\beta,\boldsymbol{\alpha};\boldsymbol{v}] \equiv \breve{\Lambda}_{\mathrm{eq}}[\beta,\boldsymbol{\alpha};\boldsymbol{v}] = S^{\mathrm{BGS}}\!\left[\hat{\breve{\varGamma}}_{\mathrm{eq}}^{[\beta,\boldsymbol{\alpha};\boldsymbol{v}]}\right] - \beta\breve{E}^{[\beta,\boldsymbol{\alpha};\boldsymbol{v}]} - \boldsymbol{\alpha}\breve{\boldsymbol{\mathcal{N}}}^{[\beta,\boldsymbol{\alpha};\boldsymbol{v}]} \tag{8}$$

will be termed *the spin Kramer fn*. For the entropy, $S^{\mathrm{BGS}}$, the Boltzmann-Gibbs-Shannon expression is used, Eq.(I.7), $S^{\mathrm{BGS}}\!\left[\hat{\varGamma}\right] = -\mathrm{Tr}\,\hat{\varGamma}\ln\hat{\varGamma}$. The first and second derivatives of $K$ w.r.t. $\{\beta,\boldsymbol{\alpha}\}$, determined according to Eqs. (I.59), (I.A13), (I.64), (∪ is here the empty set of indices, $\mathsf{L}\equiv\mathsf{F}$), are

$$K_{\mathsf{F}}[\beta,\boldsymbol{\alpha};\boldsymbol{v}] = -\left(\breve{E}^{[\beta,\boldsymbol{\alpha};\boldsymbol{v}]},\breve{\boldsymbol{\mathcal{N}}}^{[\beta,\boldsymbol{\alpha};\boldsymbol{v}]}\right), \tag{9}$$

$$K_{\mathsf{FF}}[\beta,\boldsymbol{\alpha};\boldsymbol{v}] = -\begin{pmatrix} \left(\dfrac{\partial \breve{E}^{[\beta,\boldsymbol{\alpha};\boldsymbol{v}]}}{\partial \beta}\right) & \left(\dfrac{\partial \breve{E}^{[\beta,\boldsymbol{\alpha};\boldsymbol{v}]}}{\partial \boldsymbol{\alpha}}\right) \\ \left(\dfrac{\partial \breve{\boldsymbol{\mathcal{N}}}^{[\beta,\boldsymbol{\alpha};\boldsymbol{v}]}}{\partial \beta}\right) & \left(\dfrac{\partial \breve{\boldsymbol{\mathcal{N}}}^{[\beta,\boldsymbol{\alpha};\boldsymbol{v}]}}{\partial \boldsymbol{\alpha}}\right) \end{pmatrix} = \begin{pmatrix} \langle\langle\hat{H},\hat{H}\rangle\rangle & \langle\langle\hat{H},\hat{\boldsymbol{\mathcal{N}}}^{\mathrm{T}}\rangle\rangle \\ \langle\langle\hat{\boldsymbol{\mathcal{N}}},\hat{H}\rangle\rangle & \langle\langle\hat{\boldsymbol{\mathcal{N}}},\hat{\boldsymbol{\mathcal{N}}}^{\mathrm{T}}\rangle\rangle \end{pmatrix}. \tag{10}$$

The first-order covariance matrix $\langle\langle\hat{A},\hat{B}\rangle\rangle$ is defiend in Eq.(I.3). From the symmetry of the $K_{\mathsf{FF}}$ matrix follows the identity $\left(\partial\breve{\boldsymbol{\mathcal{N}}}^{[\beta,\boldsymbol{\alpha};\boldsymbol{v}]}/\partial\beta\right) = \left(\partial\breve{E}^{[\beta,\boldsymbol{\alpha};\boldsymbol{v}]}/\partial\boldsymbol{\alpha}\right)^{\mathrm{T}}$. Note that the spin Kramer fn. $K$ is a *strictly convex* function of three variables $(\beta,\alpha_{\mathrm{N}},\alpha_{\mathrm{S}})$ and $K_{\mathsf{FF}}$ is a *positive semidefinite* (p.sd.) matrix.

The Massieu-Planck transform of the spin Kramer fn. gives *the spin grand potential* (see Eqs.(I.66), (I.68))

$$\Omega[\beta,\mu;\nu] \equiv \breve{\Upsilon}^{3,3}[\beta,\mu;\nu] = -\beta^{-1}K[\beta,-\beta\mu;\nu]$$
$$= -\beta^{-1}S^{\mathrm{BGS}}\left[\hat{\breve{\Gamma}}_{\mathrm{eq}}^{[\beta,\mu;\nu]}\right] + \breve{E}^{[\beta,\mu;\nu]} - \mu\breve{\mathcal{N}}^{[\beta,\mu;\nu]}. \tag{11}$$

Here $\mu = (\mu_{\mathrm{N}},\mu_{\mathrm{S}}) = -\beta^{-1}\alpha$. As follows from the paragraph after Eq.(4), properties of fully open $\{\hat{H},\hat{\mathcal{N}}\}$ system depend solely on the combination $(\nu(\mathbf{r})-\mu)$. Functions or functionals with the arc (in the present section) or with the tilde (in the next section) over their symbol are considered different when names of their arguments are different. Using this convention, we have $\hat{\breve{\Gamma}}_{\mathrm{eq}}[\beta,\mu;\nu] = \hat{\breve{\Gamma}}_{\mathrm{eq}}[\beta,\alpha=-\beta\mu;\nu]$, $\breve{E}[\beta,\mu;\nu] = \breve{E}[\beta,\alpha=-\beta\mu;\nu]$, etc. The Jacobian matrix, Eq.(I.A23), is

$$\breve{y}_{\mathsf{F},\mathsf{F}}^{3,3}[\beta,\mu] = \left(\begin{array}{c|c} 1 & \mathbf{0}_{\mathsf{F}'}^{\mathrm{T}} \\ \hline -\mu & -\beta\mathbf{1}_{\mathsf{F}'\mathsf{F}'} \end{array}\right) \tag{12}$$

($\mathsf{U}$ is here the empty set of indices, $\mathsf{L} \equiv \mathsf{F} = \{1,2,3\}$, $\mathsf{F}' = \{2,3\}$; $\mathbf{1}_{\mathsf{F}'\mathsf{F}'} = \{\delta_{ij}\}_{i,j\in\mathsf{F}'}$, $\mathbf{0}_{\mathsf{F}'}$ and $\mathbf{0}_{\mathsf{F}'\mathsf{F}'}$ denote zero vector and zero matrix for an appropriate index set). The first and second derivatives of $\Omega$ w.r.t. $\{\beta,\mu\}$ (using Eqs. (I.70) and (I.71)) are

$$\Omega_{\mathsf{F}}[\beta,\mu;\nu] = \left(\beta^{-2}S^{\mathrm{BGS}}, -\breve{\mathcal{N}}^{[\beta,\mu;\nu]}\right), \tag{13}$$

$$\Omega_{\mathsf{FF}}[\beta,\mu;\nu] = -\beta^{-1}\left(2\beta^{-2}S^{\mathrm{BGS}}\left(\begin{array}{c|c} 1 & \mathbf{0}_{\mathsf{F}'}^{\mathrm{T}} \\ \hline \mathbf{0}_{\mathsf{F}'} & \mathbf{0}_{\mathsf{F}'\mathsf{F}'} \end{array}\right) + \left(\breve{y}_{\mathsf{F},\mathsf{F}}^{3,3}\right)^{\mathrm{T}} K_{\mathsf{FF}}^{[\beta,-\beta\mu;\nu]}\breve{y}_{\mathsf{F},\mathsf{F}}^{3,3}\right), \tag{14}$$

with $S^{\mathrm{BGS}} = S^{\mathrm{BGS}}\left[\hat{\breve{\Gamma}}_{\mathrm{eq}}^{[\beta,\mu;\nu]}\right]$, $\breve{y}_{\mathsf{F},\mathsf{F}}^{3,3} = \breve{y}_{\mathsf{F},\mathsf{F}}^{3,3}[\beta,\mu]$. As $\beta$ means the reciprocal temperature, $\beta > 0$, the spin grand potential $\Omega$ is a *strictly concave* fn. of its arguments $\{\beta,\mu\}$ and $\Omega_{\mathsf{FF}}$ is a *negative semidefinite* (n.sd.) matrix.

The Legendre transform of the spin Kramer fn. with respect to $(\alpha_N, \alpha_S)$, defines *the spin Massieu fn.* (see Eqs. (I.42), (I.52))

$$Y[\beta, \mathcal{N}; v] \equiv \breve{\Theta}^{3,1}[\beta, \mathcal{N}; v] = K[\beta, \breve{\alpha}^{[\beta, \mathcal{N};v]}; v] + \mathcal{N}\breve{\alpha}^{[\beta, \mathcal{N};v]}$$
$$= S^{BGS}\left[\hat{\breve{\Gamma}}_{eq}^{[\beta, \mathcal{N};v]}\right] - \beta \breve{E}^{[\beta, \mathcal{N};v]}, \qquad (15)$$

where $S^{BGS}$ was substituted for $\Phi_{eq}[\breve{E}[\beta, \mathcal{N}; v], \mathcal{N}; v]$ in Eq.(I.52). Here $\breve{E}[\beta, \mathcal{N}; v] = \breve{E}[\beta, \alpha = \breve{\alpha}[\beta, \mathcal{N}; v]; v]$ (and similarly for $\hat{\breve{\Gamma}}_{eq}$), while the 2vec function $\alpha = \breve{\alpha}[\beta, \mathcal{N}; v]$ is, at fixed $\beta; v$, the reciprocal 2vec fn. to $\mathcal{N} = \breve{\mathcal{N}}[\beta, \alpha; v]$, i.e., the identities $\breve{\alpha}[\beta, \mathcal{N} = \breve{\mathcal{N}}[\beta, \alpha; v]; v] = \alpha$ and $\breve{\mathcal{N}}[\beta, \alpha = \breve{\alpha}[\beta, \mathcal{N}; v]; v] = \mathcal{N}$ hold. The spin Massieu function and the equilibrium DM can be also obtained (see Eqs. (I.61), (I.62), and (I.63)) as

$$Y[\beta, \mathcal{N}; v] = \ln\left(\text{Tr} \exp \hat{\breve{O}}^{3,1}\right), \qquad \hat{\Gamma}_{eq}[\beta, \mathcal{N}; v] = \exp \hat{\breve{O}}^{3,1} / \text{Tr} \hat{\breve{O}}^{3,1} \qquad (16)$$

from the conditions operator

$$\hat{\breve{O}}^{3,1} \equiv \hat{\breve{O}}^{3,1}[\beta, \mathcal{N}; v] = -\beta \hat{H}[v] - \breve{\alpha}^{[\beta, \mathcal{N};v]}(\hat{\mathcal{N}} - \mathcal{N}). \qquad (17)$$

First and second derivatives of $Y$ w.r.t. $\{\beta, \mathcal{N}\}$, determined according to Eqs.(I.59), (I.64),(I.65) are

$$Y_F[\beta, \mathcal{N}; v] = \left(-\breve{E}^{[\beta, \mathcal{N};v]}, \breve{\alpha}^{[\beta, \mathcal{N};v]}\right), \qquad (18)$$

$$Y_{FF}[\beta, \mathcal{N}; v] = \begin{pmatrix} -\left(\dfrac{\partial \breve{E}^{[\beta, \mathcal{N};v]}}{\partial \beta}\right) & -\left(\dfrac{\partial \breve{E}^{[\beta, \mathcal{N};v]}}{\partial \mathcal{N}}\right) \\ \left(\dfrac{\partial \breve{\alpha}^{[\beta, \mathcal{N};v]}}{\partial \beta}\right) & \left(\dfrac{\partial \breve{\alpha}^{[\beta, \mathcal{N};v]}}{\partial \mathcal{N}}\right) \end{pmatrix}$$

$$= \begin{pmatrix} \langle\langle \hat{H}, \hat{H} \rangle\rangle - \langle\langle \hat{H}, \hat{\mathcal{N}}^T \rangle\rangle \langle\langle \hat{\mathcal{N}}, \hat{\mathcal{N}}^T \rangle\rangle^{-1} \langle\langle \hat{\mathcal{N}}, \hat{H} \rangle\rangle & -\langle\langle \hat{H}, \hat{\mathcal{N}}^T \rangle\rangle \langle\langle \hat{\mathcal{N}}, \hat{\mathcal{N}}^T \rangle\rangle^{-1} \\ -\langle\langle \hat{\mathcal{N}}, \hat{\mathcal{N}}^T \rangle\rangle^{-1} \langle\langle \hat{\mathcal{N}}, \hat{H} \rangle\rangle & -\langle\langle \hat{\mathcal{N}}, \hat{\mathcal{N}}^T \rangle\rangle^{-1} \end{pmatrix}. \qquad (19)$$

From general properties of the (3,1) Massieu fn. follow: (i) $\left(\partial \breve{E}^{[\beta,\mathcal{N};v]}/\partial\beta\right)<0$ (the energy is an increasing function of the temperature, as it is expected) and (ii) $\left(\partial\breve{\alpha}^{[\beta,\mathcal{N};v]}/\partial\mathcal{N}\right)$ is n.sd. matrix (see the comment below Eq.(I.65)).

According to the discussion held in the last paragraph of Sec. III of the Paper I, the $\{\hat{H},\hat{\mathcal{N}},\hat{\mathcal{S}}\}$ system (the basis of TSDFT) can be reduced to the $\{\hat{H},\hat{\mathcal{N}}\}$ system (the basis of the spinless version of this theory) by demanding the spin Massieu fn. $Y[\beta,\mathcal{N},\mathcal{S};v]$ to reach a maximum w.r.t the spin-number variable $\mathcal{S}$. For the unique maximizer $\mathcal{S}=\mathcal{S}^0[\beta,\mathcal{N};v]$, due to $0=(\partial Y/\partial\mathcal{S})_{\mathcal{S}=\mathcal{S}^0}\equiv\breve{\alpha}_s\left[\beta,\mathcal{N},\mathcal{S}^0;v\right]$, the conditions operator $\hat{\breve{O}}^{3,1}$, Eq.(17), is reduced to the form $\hat{\breve{O}}^{3,1}\left[\beta,\mathcal{N},\mathcal{S}^0;v\right]=-\beta\hat{H}[v]-\breve{\alpha}_N\left[\beta,\mathcal{N},\mathcal{S}^0;v\right](\hat{\mathcal{N}}-\mathcal{N})$ (idependent of the operator $\hat{\mathcal{S}}$), i.e., to the conditional operator $\hat{\breve{O}}^{2,1}[\beta,\mathcal{N};v]$ that defines the Massieu fn. $Y[\beta,\mathcal{N};v]$ of the $\{\hat{H},\hat{\mathcal{N}}\}$ system. The spinless theory is relized at optimum spin-number value of the spin-dependent theory.

The Massieu-Planck transform of the spin Massieu fn. gives the *spin Helmholtz fn.* (or *spin Helmholtz free energy*)

$$A[\beta,\mathcal{N};v]\equiv\breve{\Upsilon}^{3,1}[\beta,\mathcal{N};v]=-\beta^{-1}Y[\beta,\mathcal{N};v]$$
$$=-\beta^{-1}S^{\text{BGS}}\left[\hat{\breve{\Gamma}}_{\text{eq}}^{[\beta,\mathcal{N};v]}\right]+\breve{E}^{[\beta,\mathcal{N};v]}=\Omega\left[\beta,\breve{\mu}^{[\beta,\mathcal{N};v]};v\right]+\mathcal{N}\breve{\mu}^{[\beta,\mathcal{N};v]},\qquad(20)$$

where $\breve{\mu}[\beta,\mathcal{N};v]=-\beta^{-1}\breve{\alpha}[\beta,\mathcal{N};v]$. Here $\mathsf{L}=\{1\}$, $\mathsf{U}=\{2,3\}$. The Jacobian matrix, Eq.(I.A23), is simply the unit matrix $\breve{y}_{\mathsf{F},\mathsf{F}}^{3,1}[\beta,\mathcal{N}]=1_{\mathsf{FF}}$. The first and second derivatives of $A$ w.r.t. $\{\beta,\mathcal{N}\}$, Eqs.(I.70), (I.71), are

$$A_{\mathsf{F}}[\beta,\mathcal{N};v]=\left(\beta^{-2}S^{\text{BGS}},\breve{\mu}^{[\beta,\mathcal{N};v]}\right),\qquad(21)$$

$$A_{\mathsf{FF}}[\beta,\mathcal{N};v]=-\beta^{-1}\left(\left(\begin{array}{c|c}2\beta^{-2}S^{\text{BGS}} & (\breve{\mu}[\beta,\mathcal{N};v])^{\mathrm{T}}\\\hline\breve{\mu}[\beta,\mathcal{N};v] & 0_{\mathsf{UU}}\end{array}\right)+Y_{\mathsf{FF}}\right).\qquad(22)$$

The spin Helmholtz fn. $A[\beta,\mathcal{N};\nu]$ is a *concave* fn. of $\beta$ and a *convex* fn. of $\mathcal{N}$.

Table I presents a set of the Legendre transforms $\breve{\Theta}^{3,m}$ and the conditions operators $\hat{\tilde{O}}^{3,m}$ for $m=3,2,1,0$, and the Massieu-Planck transforms $\breve{\Upsilon}^{3,m}$ for $m=3,2,1$, all at the initial ordering $\{\hat{H},\hat{\mathcal{N}},\hat{\mathcal{S}}\}$ and, for $m=2$, also at the modified ordering $\{\hat{H},\hat{\mathcal{S}},\hat{\mathcal{N}}\}\Rightarrow(j)$. Here are equations for determining the new fns. $\breve{\alpha}_S[\beta,\alpha_N,\mathcal{S};\nu]$, $\breve{\alpha}_N[\beta,\alpha_S,\mathcal{N};\nu]$, $\breve{\beta}[E,\mathcal{N};\nu]$ as their solutions

$$\breve{\mathcal{S}}[\beta,\alpha_N,\alpha_S=\breve{\alpha}_S[\beta,\alpha_N,\mathcal{S};\nu];\nu]=\mathcal{S}, \tag{23}$$

$$\breve{\mathcal{N}}[\beta,\alpha_N=\breve{\alpha}_N[\beta,\alpha_S,\mathcal{N};\nu],\alpha_S;\nu]=\mathcal{N}, \tag{24}$$

$$\breve{E}[\beta=\breve{\beta}[E,\mathcal{N};\nu],\mathcal{N};\nu]=E, \tag{25}$$

where the last equation uses the fn. defined as

$$\breve{E}[\beta,\mathcal{N};\nu]=\breve{E}[\beta,\boldsymbol{\alpha}=\breve{\boldsymbol{\alpha}}[\beta,\mathcal{N};\nu];\nu]. \tag{26}$$

The remaining new fns. are obtained using these above determined fns., e.g.,

$$\breve{\mathcal{N}}[\beta,\alpha_N,\mathcal{S};\nu]=\breve{\mathcal{N}}[\beta,\alpha_N,\alpha_S=\breve{\alpha}_S[\beta,\alpha_N,\mathcal{S};\nu];\nu], \tag{27}$$

$$\breve{\mathcal{S}}[\beta,\alpha_S,\mathcal{N};\nu]=\breve{\mathcal{S}}[\beta,\alpha_N=\breve{\alpha}_N[\beta,\alpha_S,\mathcal{N};\nu],\alpha_S;\nu], \tag{28}$$

$$\breve{\boldsymbol{\alpha}}[E,\mathcal{N};\nu]=\breve{\boldsymbol{\alpha}}[\beta=\breve{\beta}[E,\mathcal{N};\nu],\mathcal{N};\nu]. \tag{29}$$

The functions occurring in the Massieu-Planck transforms are defined by means of the substitution $\boldsymbol{\mu}=-\beta^{-1}\boldsymbol{\alpha}$, e.g.,

$$\breve{\mu}_S[\beta,\mu_N,\mathcal{S};\nu]=-\beta^{-1}\breve{\alpha}_S[\beta,\alpha_N=-\beta\mu_N,\mathcal{S};\nu], \tag{30}$$

$$\breve{\mathcal{S}}[\beta,\mu_S,\mathcal{N};\nu]=\breve{\mathcal{S}}[\beta,\alpha_S=-\beta\mu_S,\mathcal{N};\nu]. \tag{31}$$

The functions $K[\beta,\boldsymbol{\alpha};\nu]$ and $Y[\beta,\mathcal{N};\nu]$ of the $\{\hat{H},\hat{\mathcal{N}}\}$ system are known as the state finction in the entropy representation, while the function $\Omega[\beta,\boldsymbol{\mu};\nu]$ and $A[\beta,\mathcal{N};\nu]$ – in the energy representation.

## IV. EQUILIBRIUM CHARACTERISTICS OF $\{\hat{F}_{\text{int}}, \hat{\boldsymbol{\rho}}\}$ SYSTEM

The Lagrange fn. of this systems for *arbitrary* DM operator $\hat{\Gamma}$, Eq.(I.4), is

$$\tilde{\Lambda}[\hat{\Gamma}] = \Lambda[\hat{\Gamma}, \hat{\tilde{O}}] = S^{\text{BGS}}[\hat{\Gamma}] - b\,\text{Tr}\,\hat{\Gamma}\hat{F}_{\text{int}} - \int[\boldsymbol{w}\,\text{Tr}\,\hat{\Gamma}\hat{\boldsymbol{\rho}}] = S^{\text{BGS}}[\hat{\Gamma}] + \text{Tr}\,\hat{\Gamma}\hat{\tilde{O}}, \quad (32)$$

where the general conditions operator

$$\hat{\tilde{O}} \equiv \hat{\tilde{O}}[b, \boldsymbol{w}] = -b\hat{F}_{\text{int}} - \int[\boldsymbol{w}\hat{\boldsymbol{\rho}}] \quad (33)$$

depends on the Lagrange multipliers (sources) $b, \boldsymbol{w} = (w_{\text{N}}(\mathbf{r}), w_{\text{S}}(\mathbf{r}))$. The notation $\int[f] \equiv \int f(\mathbf{r})d\mathbf{r}$ is used. The dependence on the observable operators $\{\hat{F}_{\text{int}}, \hat{\boldsymbol{\rho}}\}$ is suppressed because they are universal. Since $b$ plays the role similar to the reciprocal temperature, it will be denoted traditionally $\beta$ hereafter in this section, while the individual roles of $b$ here and of $\beta$ (of Sec. III) will be clarified in Sec. V. The maximum entropy principle, Eq.(I.10),

$$\tilde{\Lambda}_{\text{eq}}[\beta, \boldsymbol{w}] = \underset{\{\hat{\Gamma}:\text{Tr}\hat{\Gamma}=1\}}{\text{Max}} \left\{ \Lambda[\hat{\Gamma}, \hat{\tilde{O}}[\beta, \boldsymbol{w}]] \right\} = \Lambda[\hat{\Gamma}_{\text{eq}}^{[\beta, \boldsymbol{w}]}, \hat{\tilde{O}}[\beta, \boldsymbol{w}]] \quad (34)$$

defines the equilibrium state of the open system via its eq-DM operator, the maximizer. In the spirit of Levy constrained-search formulation of DFT[40], the maximization in Eq.(34) will be performed in two steps: the external one w.r.t. 2vec $\boldsymbol{\rho}(\mathbf{r})$, and the internal one w.r.t. such $\hat{\Gamma}$ that it yields the given $\boldsymbol{\rho}(\mathbf{r})$:

$$\tilde{\Lambda}_{\text{eq}}[\beta, \boldsymbol{w}] = \underset{\boldsymbol{\rho}}{\text{Max}} \left\{ \underset{\{\hat{\Gamma}:\text{Tr}\hat{\Gamma}=1, \langle\hat{\boldsymbol{\rho}}\rangle_{\hat{\Gamma}}=\boldsymbol{\rho}\}}{\text{Max}} \left\{ S^{\text{BGS}}[\hat{\Gamma}] - \beta\langle\hat{F}_{\text{int}}\rangle_{\hat{\Gamma}} \right\} - \int[\boldsymbol{w}\boldsymbol{\rho}] \right\}. \quad (35)$$

The notation $\langle\hat{O}_j\rangle_{\hat{\Gamma}} = \text{Tr}\,\hat{\Gamma}\hat{O}_j$ for the expectation value of $\hat{O}_j$ is used. To have here definition of $\tilde{\Lambda}_{\text{eq}}$ meaningful, the 2vec fn. $\boldsymbol{\rho}(\mathbf{r})$ should be ensemble representable, and in addition, it should restrict the Fock space for $\hat{\Gamma} = \hat{\Gamma}[\boldsymbol{\rho}]$ (i.e., satisfying $\text{Tr}\,\hat{\Gamma}\hat{\boldsymbol{\rho}}(\mathbf{r}) = \boldsymbol{\rho}(\mathbf{r})$) to the subspace where $\operatorname{Tr} \hat{\Gamma}[\rho] \hat{F}_{\text{int}} < \infty$. It is known (see, e.g. Refs. [11,57-59]) that $\rho_\sigma(\mathbf{r})$ satisfying these condition canbe characterized by requirements: $\rho_\sigma(\mathbf{r}) \geq 0$, $\int [\rho_\sigma] < \infty$, $\int \left[ \left| \nabla \rho_\sigma^{1/2} \right|^2 \right] < \infty$. Here $\sigma \in \{\uparrow, \downarrow\}$, $\boldsymbol{\rho}(\mathbf{r}) = (\rho_\uparrow(\mathbf{r}) + \rho_\downarrow(\mathbf{r}), \rho_\uparrow(\mathbf{r}) - \rho_\downarrow(\mathbf{r}))$. The space of admissible 2vec fns. $\mathbf{w}(\mathbf{r})$ is determined by requiment $\left| \int [\boldsymbol{\rho} \mathbf{w}] \right| < \infty$ (for admissible $\boldsymbol{\rho}(\mathbf{r})$), see e.g. the space $X^*$ described in Ref. [59].

The result of the internal maximization in Eq.(35) is a universal fnl. of the density at the given $\beta$:

$$G[\beta, \boldsymbol{\rho}] = \underset{\{\hat{\Gamma}: \langle \hat{\boldsymbol{\rho}} \rangle_{\hat{\Gamma}} = \boldsymbol{\rho}\}}{\operatorname{Max}} \left\{ S^{\text{BGS}}[\hat{\Gamma}] - \beta \langle \hat{F}_{\text{int}} \rangle_{\hat{\Gamma}} \right\} = S^{\text{BGS}}\left[ \hat{\tilde{\Gamma}}_G^{[\beta, \boldsymbol{\rho}]} \right] - \beta \operatorname{Tr} \hat{\tilde{\Gamma}}_G^{[\beta, \boldsymbol{\rho}]} \hat{F}_{\text{int}} \qquad (36)$$

(its interpretation will be given in Eqs.(51) and (52), in terms of the maximizer $\hat{\tilde{\Gamma}}_G^{[\beta, \boldsymbol{\rho}]}$]. Eq.(35) can be rewritten now as

$$\tilde{\Lambda}_{\text{eq}}[\beta, \mathbf{w}] = \underset{\boldsymbol{\rho}}{\operatorname{Max}} \left\{ G[\beta, \boldsymbol{\rho}] - \int [\mathbf{w} \boldsymbol{\rho}] \right\} = G\left[ \beta, \tilde{\boldsymbol{\rho}}_G^{[\beta, \mathbf{w}]} \right] - \int \left[ \mathbf{w} \tilde{\boldsymbol{\rho}}_G^{[\beta, \mathbf{w}]} \right], \qquad (37)$$

where the maximizer $\tilde{\boldsymbol{\rho}}_G[\beta, \mathbf{w}](\mathbf{r})$ is the equilibrium density 2vec fn. of the open system at the given sources $(\beta, \mathbf{w})$. But the eq-DM at the given $(\beta, \mathbf{w})$, the maximizer in Eq.(34), is known to be

$$\hat{\tilde{\Gamma}}_{\text{eq}}[\beta, \mathbf{w}] = \exp\left( \hat{\tilde{O}}[\beta, \mathbf{w}] \right) \bigg/ \operatorname{Tr} \left( \exp\left( \hat{\tilde{O}}[\beta, \mathbf{w}] \right) \right) \qquad (38)$$

(see Eq.(I.8)). This eq-DM operator defines the characteristic fnl. (the basic Massieu functional) for the ensemble

$$\begin{aligned} X[\beta, \mathbf{w}] &\equiv \tilde{\Theta}^{3,3}[\beta, \mathbf{w}] = \tilde{\Lambda}_{\text{eq}}[\beta, \mathbf{w}] \\ &= S^{\text{BGS}}\left[ \hat{\tilde{\Gamma}}_{\text{eq}}^{[\beta, \mathbf{w}]} \right] - \beta \tilde{F}_{\text{int}}^{[\beta, \mathbf{w}]} - \int \left[ \mathbf{w} \tilde{\boldsymbol{\rho}}^{[\beta, \mathbf{w}]} \right] = G\left[ \beta, \tilde{\boldsymbol{\rho}}^{[\beta, \mathbf{w}]} \right] - \int \left[ \mathbf{w} \tilde{\boldsymbol{\rho}}^{[\beta, \mathbf{w}]} \right], \end{aligned} \qquad (39)$$

which will be termed *the spin Kramer fnl*. Here $\tilde{F}_{\text{int}}[\beta, \mathbf{w}] = \operatorname{Tr} \hat{\tilde{\Gamma}}_{\text{eq}}^{[\beta, \mathbf{w}]} \hat{F}_{\text{int}}$, $\tilde{\boldsymbol{\rho}}[\beta, \mathbf{w}](\mathbf{r}) = \operatorname{Tr} \hat{\tilde{\Gamma}}_{\text{eq}}^{[\beta, \mathbf{w}]} \hat{\boldsymbol{\rho}}(\mathbf{r})$. Since the maximum and the maximizer are unique, we have

$$\hat{\tilde{\varGamma}}_{\text{eq}}[\beta,\boldsymbol{w}] = \hat{\tilde{\varGamma}}_{\text{G}}\big[\beta,\tilde{\boldsymbol{\rho}}_{\text{G}}[\beta,\boldsymbol{w}]\big], \qquad \tilde{\boldsymbol{\rho}}[\beta,\boldsymbol{w}](\mathbf{r}) = \tilde{\boldsymbol{\rho}}_{\text{G}}[\beta,\boldsymbol{w}](\mathbf{r}), \tag{40}$$

and the last form of $X[\beta,\boldsymbol{w}]$ in Eq.(39). Using Eqs. (I.60), (I.65), (∪ is here the empty set, L ≡ F), the first and second derivatives of $X$ are

$$X_{\text{F}}[\beta,\boldsymbol{w}](\mathbf{r}) = \left(\left(\frac{\partial X[\beta,\boldsymbol{w}]}{\partial \beta}\right), \left(\frac{\delta X[\beta,\boldsymbol{w}]}{\delta \boldsymbol{w}(\mathbf{r})}\right)^{\text{T}}\right) = -\left(F_{\text{int}}^{[\beta,\boldsymbol{w}]}, \boldsymbol{\rho}^{[\beta,\boldsymbol{w}]}(\mathbf{r})\right), \tag{41}$$

$$X_{\text{FF}}[\beta,\boldsymbol{w}](\mathbf{r},\mathbf{r}') = -\left(\begin{array}{c|c} \left(\dfrac{\partial F_{\text{int}}^{[\beta,\boldsymbol{w}]}}{\partial \beta}\right) & \left(\dfrac{\delta F_{\text{int}}^{[\beta,\boldsymbol{w}]}}{\delta \boldsymbol{w}(\mathbf{r}')}\right) \\ \hline \left(\dfrac{\partial \boldsymbol{\rho}^{[\beta,\boldsymbol{w}]}(\mathbf{r})}{\partial \beta}\right) & \left(\dfrac{\delta \boldsymbol{\rho}^{[\beta,\boldsymbol{w}]}(\mathbf{r})}{\delta \boldsymbol{w}(\mathbf{r}')}\right) \end{array}\right)$$
$$= \left(\begin{array}{c|c} \langle\langle \hat{F}_{\text{int}}, \hat{F}_{\text{int}} \rangle\rangle & \langle\langle \hat{F}_{\text{int}}, \hat{\boldsymbol{\rho}}^{\text{T}}(\mathbf{r}') \rangle\rangle \\ \hline \langle\langle \hat{\boldsymbol{\rho}}(\mathbf{r}), \hat{F}_{\text{int}} \rangle\rangle & \langle\langle \hat{\boldsymbol{\rho}}(\mathbf{r}), \hat{\boldsymbol{\rho}}^{\text{T}}(\mathbf{r}') \rangle\rangle \end{array}\right). \tag{42}$$

(A partial derivative w.r.t. a local (**r**-dependent) quantity is, in fact, a fnl. derivative; therefore, the second derivatives represent a 3×3 matrix and, simultaneously, a $(\mathbf{r},\mathbf{r}')$-dependent kernel). Similarly as for the $\{\hat{H}, \hat{\mathcal{N}}\}$ system, $X_{\text{FF}}$ is a p.sd. matrix-kernel, and this means that $\left(\partial F_{\text{int}}^{[\beta,\boldsymbol{w}]}/\partial \beta\right) < 0$ (the internal energy is an increasing function of the temperature, as it is expected) and $\left(\delta \boldsymbol{\rho}^{[\beta,\boldsymbol{w}]}(\mathbf{r})/\delta \boldsymbol{w}(\mathbf{r}')\right)$ is a n.sd. matrix-kernel.

*The spin grand functional* is the Massieu-Planck transform of the spin Kramer fnl.

$$B[\beta,\boldsymbol{u}] \equiv \tilde{\Upsilon}^{3,3}[\beta,\boldsymbol{u}] = -\beta^{-1} X[\beta,\boldsymbol{w} = -\beta\boldsymbol{u}] = -\beta^{-1} G\big[\beta,\tilde{\boldsymbol{\rho}}^{[\beta,\boldsymbol{u}]}\big] - \int\big[\boldsymbol{u}\tilde{\boldsymbol{\rho}}^{[\beta,\boldsymbol{u}]}\big] \tag{43}$$

where $\boldsymbol{u}(\mathbf{r}) = -\beta^{-1}\boldsymbol{w}(\mathbf{r})$ and $\tilde{\boldsymbol{\rho}}[\beta,\boldsymbol{u}] = \tilde{\boldsymbol{\rho}}[\beta,\boldsymbol{w} = -\beta\boldsymbol{u}]$. The Jacobian matrix fnl., Eq.(I.A23) (∪ is here the empty set, L ≡ F), is

$$\tilde{y}_{F,F}^{3,3}[\beta,u](\mathbf{r},\mathbf{r}') = \begin{pmatrix} \left(\dfrac{\partial \beta}{\partial \beta}\right) & \left(\dfrac{\delta \beta}{\delta u(\mathbf{r}')}\right) \\ \hline \left(\dfrac{\partial w(\mathbf{r})}{\partial \beta}\right) & \left(\dfrac{\delta w(\mathbf{r})}{\delta u(\mathbf{r}')}\right) \end{pmatrix} = \begin{pmatrix} 1 & 0_{F'}^{T} \\ \hline -u(\mathbf{r}) & -\beta\delta(\mathbf{r}-\mathbf{r}')1_{F'F'} \end{pmatrix}. \quad (44)$$

Using Eqs. (I.70) and (I.71), the first and second derivatives are

$$B_{F}[\beta,u](\mathbf{r}) = \left(\left(\dfrac{\partial B[\beta,u]}{\partial \beta}\right),\left(\dfrac{\delta B[\beta,u]}{\delta u(\mathbf{r})}\right)^{T}\right) = \left(\beta^{-2}S^{BGS},-\tilde{\rho}^{[\beta,u]}(\mathbf{r})\right), \quad (45)$$

$$B_{FF}[\beta,u](\mathbf{r},\mathbf{r}') = -\beta^{-1}\left(2\beta^{-2}S^{BGS}\begin{pmatrix}1 & 0_{F'}^{T} \\ \hline 0_{F'} & 0_{F'F'}\end{pmatrix} + \left(\tilde{y}_{F,F}^{3,3}\right)^{T}X_{FF}\tilde{y}_{F,F}^{3,3}\right). \quad (46)$$

The matrix multiplication includes here also an integration over a dummy spatial variable. We give explicit elements of this second derivative matrix-kernel, Eq.(46):

$$\dfrac{\partial^{2}B[\beta,u]}{\partial \beta^{2}} = -\beta^{-1}\left\{2\beta^{-2}S^{BGS} + \dfrac{\partial^{2}X[\beta,w]}{\partial \beta^{2}} - 2\int\dfrac{\partial \delta X[\beta,w]}{\partial \beta \delta w(\mathbf{r})}u(\mathbf{r})d\mathbf{r} \right.$$
$$\left. + \int u^{T}(\mathbf{r})\dfrac{\delta^{2}X[\beta,w]}{\delta w(\mathbf{r})\delta w(\mathbf{r}')}u(\mathbf{r}')d\mathbf{r}d\mathbf{r}'\right\}, \quad (47)$$

$$\dfrac{\partial}{\partial \beta}\left(\dfrac{\delta B[\beta,u]}{\delta u(\mathbf{r})}\right)^{T} = \dfrac{\partial}{\partial \beta}\left(\dfrac{\delta X[\beta,w]}{\delta w(\mathbf{r})}\right)^{T} - \int\dfrac{\delta^{2}X[\beta,w]}{\delta w(\mathbf{r})\delta w(\mathbf{r}')}u(\mathbf{r}')d\mathbf{r}', \quad (48)$$

$$\dfrac{\delta^{2}B[\beta,u]}{\delta u(\mathbf{r})\delta u(\mathbf{r}')} = -\beta\dfrac{\delta^{2}X[\beta,w]}{\delta w(\mathbf{r})\delta w(\mathbf{r}')}, \quad (49)$$

where, after performing differentiations of $X$, its argument $w(\mathbf{r})$ should be replaced by $-\beta u(\mathbf{r})$ in the result. According to the comment bellow Eq.(I.77), when $\beta > 0$, the matrix-kernel $B_{FF}$ is a n.sd. one, and therefore as discussed in Appendix B of Paper I, $B_{1,1} < 0$, and $B_{F'F'}$ is also a n.sd. matrix-kernel.

*The spin Massieu universal fnl.* is the Legendre transform of the spin Kramer fnl. $X$ w.r.t. $w(\mathbf{r}) = \tilde{w}[\beta,\rho](\mathbf{r})$:

$$\tilde{\Theta}^{3,1}[\beta,\rho] = \tilde{\Theta}^{3,3}[\beta, w = \tilde{w}[\beta,\rho]] + \int [\rho \tilde{w}^{[\beta,\rho]}] = S^{\text{BGS}}\left[\hat{\tilde{\Gamma}}_{\text{eq}}^{[\beta,\rho]}\right] - \beta \tilde{F}_{\text{int}}[\beta,\rho], \qquad (50)$$

where $\tilde{F}_{\text{int}}[\beta,\rho] = \tilde{F}_{\text{int}}[\beta, w = \tilde{w}[\beta,\rho]]$, the 2vec fn. $w(\mathbf{r}) = \tilde{w}[\beta,\rho](\mathbf{r})$ is, at fixed $\beta$, the reciprocal 2vec fn. to $\rho(\mathbf{r}) = \tilde{\rho}[\beta, w](\mathbf{r})$, i.e., the identities $\tilde{w}[\beta, \rho = \tilde{\rho}[\beta, w]](\mathbf{r}) = w(\mathbf{r})$ and $\tilde{\rho}[\beta, w = \tilde{w}[\beta,\rho]](\mathbf{r}) = \rho(\mathbf{r})$ hold. Due to the last identity applied to $X \equiv \tilde{\Theta}^{3,3}$ in terms of $G$, Eq.(39), the fnl. $\tilde{\Theta}^{3,1}[\beta,\rho]$, Eq.(50), can be written also as

$$\tilde{\Theta}^{3,1}[\beta,\rho] = G[\beta,\rho] = S^{\text{BGS}}\left[\hat{\tilde{\Gamma}}_{\text{eq}}^{[\beta,\rho]}\right] - \beta \tilde{F}_{\text{int}}^{[\beta,\rho]}. \qquad (51)$$

We see that $G[\beta,\rho]$ defined with maximization in Eq.(36) is, in fact, the spin *Massieu universal fnl*. This fnl. can be also obtained as

$$G[\beta,\rho] \equiv \tilde{\Theta}^{3,1}[\beta,\rho] = \ln\left(\text{Tr}\exp\left(\hat{\tilde{O}}^{3,1}[\beta,\rho]\right)\right) \qquad (52)$$

from the conditions operator, Eq.(I.61),

$$\hat{\tilde{O}}^{3,1}[\beta,\rho] = -\beta \hat{F}_{\text{int}} - \int\left[\tilde{w}^{[\beta,\rho]}(\hat{\rho} - \rho)\right]. \qquad (53)$$

According to Eqs. (I.59) and (I.64), the first and second derivatives of $G[\beta,\rho]$ are

$$G_{\text{F}}[\beta,\rho](\mathbf{r}) = \left(\left(\frac{\partial G[\beta,\rho]}{\partial \beta}\right), \left(\frac{\delta G[\beta,\rho]}{\delta \rho(\mathbf{r})}\right)^{\text{T}}\right) = \left(-F_{\text{int}}^{[\beta,\rho]}, \tilde{w}^{[\beta,\rho]}(\mathbf{r})\right), \qquad (54)$$

$$G_{\text{FF}}[\beta,\rho](\mathbf{r},\mathbf{r}') = \begin{pmatrix} -\left(\dfrac{\partial F_{\text{int}}^{[\beta,\rho]}}{\partial \beta}\right) & -\left(\dfrac{\delta F_{\text{int}}^{[\beta,\rho]}}{\delta \rho(\mathbf{r}')}\right) \\ \hdashline \left(\dfrac{\partial \tilde{w}^{[\beta,\rho]}(\mathbf{r})}{\partial \beta}\right) & \left(\dfrac{\delta \tilde{w}^{[\beta,\rho]}(\mathbf{r})}{\delta \rho(\mathbf{r}')}\right) \end{pmatrix}$$

$$= \begin{pmatrix} \langle\langle \hat{F}_{\text{int}}, \hat{F}_{\text{int}} \rangle\rangle - \langle\langle \hat{F}_{\text{int}}, \hat{\rho}^{\text{T}}(\mathbf{r}) \rangle\rangle \langle\langle \hat{\rho}(\mathbf{r}), \hat{\rho}^{\text{T}}(\mathbf{r}') \rangle\rangle^{-1} \langle\langle \hat{\rho}(\mathbf{r}'), \hat{F}_{\text{int}} \rangle\rangle & -\langle\langle \hat{F}_{\text{int}}, \hat{\rho}^{\text{T}}(\mathbf{r}) \rangle\rangle \langle\langle \hat{\rho}(\mathbf{r}), \hat{\rho}^{\text{T}}(\mathbf{r}') \rangle\rangle^{-1} \\ \hdashline -\langle\langle \hat{\rho}(\mathbf{r}), \hat{\rho}^{\text{T}}(\mathbf{r}') \rangle\rangle^{-1} \langle\langle \hat{\rho}(\mathbf{r}'), \hat{F}_{\text{int}} \rangle\rangle & -\langle\langle \hat{\rho}(\mathbf{r}), \hat{\rho}^{\text{T}}(\mathbf{r}') \rangle\rangle^{-1} \end{pmatrix}.$$

$$(55)$$

The element $G_{11}$ is positive, while $\left(\delta \tilde{w}[\beta,\rho](\mathbf{r}) / \delta \rho(\mathbf{r}')\right)$ is a n.sd. matrix-kernel.

Similarly as in the discussion after Eq.(19), by demanding the spin Massieu universal fnl. $G[\beta,\rho_N,\rho_S]$ to reach a maximum w.r.t spin-density function $\rho_S(\mathbf{r})$, the $\{\hat{F}_{int},\hat{\rho}_N,\hat{\rho}_S\}$ system can be reduced to the $\{\hat{F}_{int},\hat{\rho}_N\}$ system. The spinless theory is realized at optimum spin-density fn. of the spin-dependent theory.

The Massieu-Planck transform of the spin Massieu universal fnl. $G[\beta,\rho]$ gives *the spin canonical universal fnl.*

$$F[\beta,\rho] \equiv \tilde{\Upsilon}^{3,1}[\beta,\rho] = -\beta^{-1}\tilde{\Theta}^{3,1}[\beta,\rho] = -\beta^{-1}G[\beta,\rho]. \tag{56}$$

Note that according to Eq.(36), for $\beta > 0$ the fnl. $F[\beta,\rho]$ can be defined as the result of minimization

$$\begin{aligned}
F[\beta,\rho] &= \underset{\{\hat{\Gamma}:\langle\hat{\rho}\rangle_{\hat{\Gamma}}=\rho\}}{\text{Min}} \left\{ \langle\hat{F}_{int}\rangle_{\hat{\Gamma}} - \beta^{-1}S^{BGS}[\hat{\Gamma}] \right\} = \text{Tr}\,\hat{\tilde{\Gamma}}_G^{[\beta,\rho]}\hat{F}_{int} - \beta^{-1}S^{BGS}\left[\hat{\tilde{\Gamma}}_G^{[\beta,\rho]}\right] \\
&= \tilde{F}_{int}[\beta,\rho] - \beta^{-1}S^{BGS}\left[\hat{\tilde{\Gamma}}_{eq}^{[\beta,\rho]}\right],
\end{aligned} \tag{57}$$

(Eq.(40) was applied for the last form of $F$). Therefore, $\lim_{\beta\to\infty} F[\beta,\rho] = F_L[\rho]$ — the Lieb-Valone fnl. of the density $\rho(\mathbf{r}) = (\rho_N(\mathbf{r}),\rho_S(\mathbf{r}))$, Eq.(A4), or $F_L[\rho]$ fnl. of the density $\rho(\mathbf{r})$, Eq.(A1), when restricted to the standard zero-temperature SDFT or DFT, see Appendix.

The Jacobian matrix-kernel is the identity matrix-kernel

$$\tilde{y}_{F,F}^{3,1}[\beta,\rho](\mathbf{r},\mathbf{r}') = \begin{pmatrix} 1 & 0_U^T \\ 0_U & \delta(\mathbf{r}-\mathbf{r}')1_{UU} \end{pmatrix}. \tag{58}$$

The first and second derivatives of $F[\beta,\rho]$, determined according to Eqs. (I.70) and (I.71), are

$$F_F[\beta,\rho](\mathbf{r}) = \left(\beta^{-2}S^{BGS}, \tilde{u}^{[\beta,\rho]}(\mathbf{r})\right), \tag{59}$$

$$F_{FF}[\beta,\rho](\mathbf{r},\mathbf{r}') = -\beta^{-1}\left(\begin{pmatrix} 2\beta^{-2}S^{BGS} & \left(\tilde{u}^{[\beta,\rho]}(\mathbf{r}')\right)^T \\ \tilde{u}^{[\beta,\rho]}(\mathbf{r}) & 0_{F'F'} \end{pmatrix} + G_{FF}(\mathbf{r},\mathbf{r}')\right), \tag{60}$$

where $\tilde{u}^{[\beta,\rho]}(\mathbf{r}) = -\beta^{-1}\tilde{w}^{[\beta,\rho]}(\mathbf{r})$. According to the comment after Eq.(I.77), for $\beta > 0$ we have $F_{11}[\tilde{\beta},\rho] < 0$ and the matrix-kernel $F_{UU} = (\delta\tilde{u}[\beta,\rho](\mathbf{r})/\delta\rho(\mathbf{r}'))$ is a p.sd. one.

Table II presents a set of the Legendre transforms $\tilde{\Theta}^{3,m}$ and the conditions operators $\hat{\tilde{O}}^{3,m}$ for $m = 3, 2, 1, 0$, and the Massieu-Planck transforms $\tilde{\Upsilon}^{3,m}$ for $m = 3, 2, 1$, all at the initial ordering $\{\hat{F}_{\text{int}}, \hat{\rho}_N, \hat{\rho}_S\}$ and, for $m = 2$, also at the modified ordering $\{\hat{F}_{\text{int}}, \hat{\rho}_S, \hat{\rho}_N\} \Rightarrow (j)$. The new fnls. $\tilde{w}_S[\beta, w_N, \rho_S]$, $\tilde{w}_N[\beta, w_S, \rho_N]$, $\tilde{\beta}[F_{\text{int}}, \rho]$ are solutions of equations

$$\tilde{\rho}_S\big[\beta, w_N, w_S = \tilde{w}_S[\beta, w_N, \rho_S]\big](\mathbf{r}) = \rho_S(\mathbf{r}), \tag{61}$$

$$\tilde{\rho}_N\big[\beta, w_N = \tilde{w}_N[\beta, w_S, \rho_N], w_S\big](\mathbf{r}) = \rho_N(\mathbf{r}), \tag{62}$$

$$\tilde{F}_{\text{int}}\big[\beta = \tilde{\beta}[F_{\text{int}}, \rho], \rho\big] = F_{\text{int}}, \tag{63}$$

where the last equation uses the fnl. defined as

$$\tilde{F}_{\text{int}}[\beta, \rho] = \tilde{F}_{\text{int}}\big[\beta, w = \tilde{w}[\beta, \rho]\big]. \tag{64}$$

The remaining new functionals are obtained using these above determined fnls., e.g.,

$$\tilde{\rho}_N[\beta, w_N, \rho_S](\mathbf{r}) = \tilde{\rho}_N\big[\beta, w_N, w_S = \tilde{w}_S[\beta, w_N, \rho_S]\big](\mathbf{r}), \tag{65}$$

$$\tilde{\rho}_S[\beta, w_S, \rho_N](\mathbf{r}) = \tilde{\rho}_S\big[\beta, w_N = \tilde{w}_N[\beta, w_S, \rho_N], w_S\big](\mathbf{r}), \tag{66}$$

$$\tilde{w}[F_{\text{int}}, \rho](\mathbf{r}) = \tilde{w}\big[\beta = \tilde{\beta}[F_{\text{int}}, \rho], \rho\big](\mathbf{r}). \tag{67}$$

The fnls. occurring in the Massieu-Planck transforms are defined by means of the substitution $u = -\beta^{-1}w$, e.g.

$$\tilde{u}_S[\beta, u_N, \rho_S](\mathbf{r}) = -\beta^{-1}\tilde{w}_S[\beta, w_N = -\beta u_N, \rho_S](\mathbf{r}), \tag{68}$$

$$\tilde{\rho}_S[\beta, u_S, \rho_N](\mathbf{r}) = \tilde{\rho}_S[\beta, w_S = -\beta u_S, \rho_N](\mathbf{r}). \tag{69}$$

The functionals $X[\beta,w]$ and $G[\beta,\rho]$ are known as the state functions in the entropy representation of the $\{\hat{F}_{int},\hat{\rho}\}$ system, while the functionals $B[\beta,u]$ and $F[\beta,\rho]$ – in the energy representation.

## V. EQUIVALENCE OF TWO SYSTEMS

While the two considered systems were defined independently, we demand now them to be equivalent, i.e. to have the same eq-DM, $\hat{\breve{\varGamma}}_{eq} = \hat{\tilde{\varGamma}}_{eq}$. Since the eq-DM is uniquely determined by the general conditions operator $\hat{O}$, Eq.(I.8), we need to demand

$$\hat{\breve{O}}^{3,3}[\beta,\alpha;v] = \hat{\tilde{O}}^{3,3}[b,w]. \tag{70}$$

To achieve this, we transform $\hat{\breve{O}}^{3,3}$, Eq.(4), to be expressed in terms of the operators $\{\hat{F}_{int},\hat{\rho}\}$ of the second system

$$\begin{aligned}\breve{O}^{3,3}[\beta,\alpha;v] &= -\beta\hat{H}[v] - \alpha\hat{\mathcal{N}} = -\beta\left(\hat{T}+\hat{U}+\int[v\hat{\rho}]\right) - \alpha\int[\hat{\rho}]\\ &= -\beta\hat{F}_{int} - \int\left[(\beta v + \alpha)\hat{\rho}\right]\end{aligned} \tag{71}$$

(using the notation $\int[f] = \int f(\mathbf{r})d\mathbf{r}$ and the identity $\int[\hat{\rho}] = \hat{\mathcal{N}}$). By comparing this expression with $\hat{\tilde{O}}^{3,3}$, Eq.(33), we find the following relations between $\{\beta,\alpha;v\}$ and $\{b,w\}$:

$$b = \beta, \qquad w(\mathbf{r}) = \beta v(\mathbf{r}) + \alpha, \tag{72}$$

guaranteeing the equivalence of two systems. Using Eq.(I.11), from the equality (70) follows $\breve{\Lambda}_{eq} = \tilde{\Lambda}_{eq}$, therefore, applying Eqs.(8) and (39), we find the relation between the spin Kramer fn. $K$ and the spin Kramer fnl. $X$

$$K[\beta,\alpha;v] = X[\beta,\beta v + \alpha]. \tag{73}$$

Since the fn. $K$ represent a property of an open $\{\hat{H}, \hat{\mathcal{N}}\}$ system, Eq.(73) confirms role of the combination $(v + \beta^{-1}\alpha) = (\beta v + \alpha)\beta^{-1}$, indicated in the paragraph after Eq.(4). So The Kramer fn. can be transformed, e.g., as

$$K[\beta, \alpha; v] = K[\beta, 0; v + \beta^{-1}\alpha] = K[\beta, \alpha + \beta v; 0]. \tag{74}$$

After expressing $X$ in terms of $G$, Eq.(39), we rewrite Eq.(73) as

$$K[\beta, \alpha; v] = G[\beta, \tilde{\rho}^{[\beta, w]}] - \int [w\tilde{\rho}^{[\beta, w]}], \tag{75}$$

where $w(\mathbf{r}) = w[\beta, \alpha; v](\mathbf{r})$ is given in Eq.(72). The Massieu fnl. written in terms of the Kramer fnl.

$$G[\beta, \rho] = K\left[\beta, (\tilde{w}^{[\beta, \rho]} - \beta v); v\right] + \int [\rho \tilde{w}^{[\beta, \rho]}] = K[\beta, 0; \beta^{-1}\tilde{w}^{[\beta, \rho]}] + \int [\rho \tilde{w}^{[\beta, \rho]}], \tag{76}$$

follows from Eqs.(73) and (75) after using the function $\tilde{w}[\beta, \rho](\mathbf{r})$ and related identities [defined and shown below Eq.(50)] and applying Eq.(72) to substitute $\alpha = \tilde{w}(\mathbf{r}) - \beta v(\mathbf{r})$ for the second argument of $K$. Next, since $b = \beta$ and $\tilde{\Lambda}_{eq} = \breve{\Lambda}_{eq}$, the Massieu-Planck transforms are identical, $\tilde{\Upsilon}^{3,3} = \breve{\Upsilon}^{3,3}$. Therefore, using Eqs.(11), (43), we express the spin grand potential $\Omega$ in terms of the spin grand fnl. $B$

$$\Omega[\beta, \mu; v] = B[\beta, \mu - v], \tag{77}$$

where the second argument of $B$,

$$u(\mathbf{r}) = -(v(\mathbf{r}) - \mu), \tag{78}$$

was obtained from Eq.(72) using definitions $\mu = -\beta^{-1}\alpha$, $u(\mathbf{r}) = -\beta^{-1}w(\mathbf{r})$. As we see, $\Omega$ depends effectively on two arguments (as was already noticed after Eq.(11)), so it can be transformed, e.g., as

$$\Omega[\beta, \mu; v] = \Omega[\beta, 0; v - \mu] = \Omega[\beta, \mu - v; 0]. \tag{79}$$

After expressing $B$ in terms of $\beta^{-1}G$, Eq.(43), i.e. $F$, Eq.(56), we rewrite Eq.(77) as

$$\Omega[\beta,0;-u] = F\left[\beta,\tilde{\rho}^{[\beta,u]}\right] - \int\left[u\tilde{\rho}^{[\beta,u]}\right], \tag{80}$$

where $\tilde{\rho}[\beta,u] = \tilde{\rho}[\beta,w=-\beta u]$. This Eq.(80) can be rearranged as

$$F[\beta,\rho] = \Omega\left[\beta,0;-\tilde{u}^{[\beta,\rho]}\right] + \int\left[\rho\tilde{u}^{[\beta,\rho]}\right], \tag{81}$$

where $\tilde{u}[\beta,\rho](\mathbf{r}) = -\beta^{-1}\tilde{w}[\beta,\rho](\mathbf{r})$, see the definition of $\tilde{w}$ below Eq.(50), so identities $\tilde{u}[\beta,\rho=\tilde{\rho}[\beta,u]](\mathbf{r}) = u(\mathbf{r})$, $\tilde{\rho}[\beta,u=\tilde{u}[\beta,\rho]](\mathbf{r}) = \rho(\mathbf{r})$ has been applied.

While two Massieu-Planck transforms, $\breve{\Upsilon}^{3,1} \equiv A[\beta,\mathcal{N};v]$ for the $\{\hat{H},\hat{\mathcal{N}}\}$ system, and $\tilde{\Upsilon}^{3,1} \equiv F[\beta,\rho]$ for the $\{\hat{F}_{int},\hat{\rho}(\mathbf{r})\}$ system, are not equal (even if two systems are equivalent), nevertheless a relation between them can be found at the equivalence conditions (72). The expectation value of the Hamiltonian, Eq.(6), can be split according to the second splitting of $\hat{H}$, Eq.(1):

$$\breve{E}\left[\beta,\mathcal{N}=\int[\rho];v\right] = \breve{E}\left[\beta,\alpha=\breve{\alpha}\left[\beta,\mathcal{N}=\int[\rho];v\right];v\right] = \langle\hat{F}_{int}\rangle_{\hat{\tilde{\Gamma}}_{eq}} + \int[v\rho], \tag{82}$$

(see the definitions below Eq.(15)), therefore the Helmholtz free energy $A$, Eq.(20), can be expressed in terms of the spin canonical fnl. $F[\beta,\rho]$, Eq.(57), using the equivalence condition $\hat{\breve{\Gamma}}_{eq}^{[\beta,\mathcal{N};v]} = \hat{\tilde{\Gamma}}_{eq}^{[\beta,\rho]}$ at $\mathcal{N} = \int[\rho]$, as

$$A\left[\beta,\int[\rho];v\right] = F[\beta,\rho] + \int[v\rho]. \tag{83}$$

From the relation (83) multiplied by $(-\beta)$ follows immediately the relation between $\breve{\Theta}^{3,1} \equiv Y$, see Eq.(20), and $\tilde{\Theta}^{3,1} \equiv G$, see Eq.(56),

$$Y\left[\beta,\int[\rho];v\right] = G[\beta,\rho] - \beta\int[v\rho]. \tag{84}$$

Table III collects conditions (like Eq.(72) and Eq.(78)) for the equivalence between two systems for various Legendre and Massieu-Planck transforms.

## VI. THE HOHENBERG-KOHN THEOREMS IN TSDFT

In this section, we discuss the extension of the Hohenberg-Kohn theorem to TSDFT. In the prior formulation of the spin extension of the Mermin theory[23] presented by Rajagopal,[24] only one chemical potential $\mu = \mu_N$ (Lagrange multiplier conjugate to $\hat{\mathcal{N}}$) occurred. This means, in our notation, that $\breve{\mu}[\beta, \mathcal{N}, \mathcal{S}] = (\breve{\mu}_N, \breve{\mu}_S = 0) = -\beta^{-1}(\breve{\alpha}_N, \breve{\alpha}_S = 0)$ was used, so the system optimized w.r.t. the spin number $\mathcal{S}$ at given $\beta$ and $\mathcal{N}$ was tacitly assumed (see the end of Sec. III of Paper I). In contrast to this formulation, in the present extension two components of $\boldsymbol{\mu} = (\mu_N, \mu_S)$ are treated as independent parameters (the one chemical potential vs. two chemical potentials issue is also discussed at zero temperature in Ref.[42]). In consequence, in the spin canonical ensemble approach presented here, two items in the pair $(\mathcal{N}, \mathcal{S})$ or $(\rho_N, \rho_S)$ are independent variables.

Using Eq.(37) with Eq.(39), we can write

$$\operatorname*{Max}_{\boldsymbol{\rho}}\{G[\beta, \boldsymbol{\rho}] - \int[\boldsymbol{w}\boldsymbol{\rho}]\} = G[\beta, \boldsymbol{\rho}_0] - \int[\boldsymbol{w}\boldsymbol{\rho}_0] = X[\beta, \boldsymbol{w}]. \tag{85}$$

Here the maximizer $\boldsymbol{\rho}_0(\mathbf{r}) = \tilde{\boldsymbol{\rho}}[\beta, \boldsymbol{w}](\mathbf{r})$ is the equilibrium density of the $\{\hat{F}_{int}, \hat{\boldsymbol{\rho}}(\mathbf{r})\}$ system at given sources $\{\beta, \boldsymbol{w}(\mathbf{r})\}$. For given sources $\{\beta, \boldsymbol{\alpha}\}$ of the equivalent $\{\hat{H}, \hat{\mathcal{N}}\}$ system, its 2vec potential $\boldsymbol{v}(\mathbf{r})$ is related to $\boldsymbol{w}(\mathbf{r})$ by means of Eq.(72). After inserting this form of $\boldsymbol{w}(\mathbf{r})$, Eq.(85) can be rewritten as

$$\begin{aligned}\operatorname*{Max}_{\boldsymbol{\rho}}\{G[\beta, \boldsymbol{\rho}] - \int[(\beta\boldsymbol{v} + \boldsymbol{\alpha})\boldsymbol{\rho}]\} &= G[\beta, \boldsymbol{\rho}_0] - \beta\int[\boldsymbol{v}\boldsymbol{\rho}_0] - \boldsymbol{\alpha}\mathcal{N}[\beta, \boldsymbol{\alpha}; \boldsymbol{v}] \\ &= K[\beta, \boldsymbol{\alpha}; \boldsymbol{v}]\end{aligned}, \tag{86}$$

where the maximizer is $\rho_0(\mathbf{r}) = \tilde{\rho}[\beta, w = \beta v + \alpha](\mathbf{r})$. For the last equality in Eq.(86), Eq.(73) with Eq.(85) is applied. We see that $\rho_0(\mathbf{r})$ represents also the equilibrium density of the $\{\hat{H}, \hat{\mathcal{N}}\}$ system. Next, let us perform the following minimization w.r.t. $w(\mathbf{r})$:

$$\underset{w}{\text{Min}}\{X[\beta, w] + \int[w\rho]\} = X[\beta, w_0] + \int[w_0\rho] = G[\beta, \rho]. \tag{87}$$

The minimizer $w_0(\mathbf{r}) = \tilde{w}[\beta, \rho](\mathbf{r})$ is the first derivative of $G[\beta, \rho]$ with respect to densities, Eq.(54). Since the spin Massieu universal functional, $G[\beta, \rho]$, is the Legendre transform of the spin Kramer functional, $X[\beta, w]$, the $w_0 \leftrightarrow \rho_0$ mapping is true.

Similar results can be obtained for the Gibbs-Helmholtz fns. of Paper I, i.e., in the energy representation. Using the definition of $F$ in Eq.(56) and $u(\mathbf{r}) = -\beta^{-1} w(\mathbf{r})$, we have

$$\underset{\rho}{\text{Min}}\{F[\beta, \rho] - \int[u\rho]\} = F[\beta, \rho_0] - \int[u\rho_0] = B[\beta, u] = -\beta^{-1} X[\beta, -\beta u]. \tag{88}$$

Here $\rho_0(\mathbf{r})$, the previous maximizer of $\{G[\beta, \rho_0] - \beta\int[v\rho_0]\}$, is now the minimizer of $\{F[\beta, \rho] - \int[u\rho]\}$, $\rho_0(\mathbf{r}) = \tilde{\rho}[\beta, w = -\beta u](\mathbf{r}) = \tilde{\rho}[\beta, u](\mathbf{r})$. Using Eq.(77) and connecting this result with the $\{\hat{H}, \hat{\mathcal{N}}\}$ system characterized by sources $\beta, \mu = -\beta^{-1}\alpha$, where $v(\mathbf{r})$ and $u(\mathbf{r})$ are related in Eq.(78), we rewrite Eq.(88) as

$$\underset{\rho}{\text{Min}}\{F[\beta, \rho] - \int[(\mu - v)\rho]\} = B[\beta, \mu - v] = \Omega[\beta, \mu; v]. \tag{89}$$

Now, the minimization in Eq.(87) can be rewritten as the maximization in the energy representation

$$\underset{u}{\text{Max}}\{B[\beta, u] + \int[u\rho]\} = B[\beta, u_0] + \int[u_0\rho] = F[\beta, \rho]. \tag{90}$$

Here the maximizer $u_0(\mathbf{r}) = \tilde{u}[\beta, \rho](\mathbf{r})$ is the first derivative of $F[\beta, \rho]$ w.r.t. density $\rho(\mathbf{r})$, Eq.(59). Since the spin-canonical universal functional, $F[\beta, \rho]$, is the Legendre transform of the spin-grand functional, $B[\beta, u]$, the $u_0 \leftrightarrow \rho_0$ mapping is true.

These results can be collected in the form of theorems. In the entropy representation (i.e. using Massieu fns. of Paper I):

**Theorem 1.** *In the spin-grand-canonical ensemble at a given inverse temperature $\beta$ and in contact with a reservoir of the electrons characterized by a 2vec $\boldsymbol{\alpha}$, the 2vec equilibrium electron density $\rho_0(\mathbf{r})$ is determined uniquely by the dimensionless potential $w(\mathbf{r}) = (\beta v(\mathbf{r}) + \boldsymbol{\alpha})$. This density $\rho_0$ maximizes w.r.t. $\rho$ the $\{G[\beta, \rho] - \int [w\rho]\}$ functional at fixed $\beta$ and fixed the 2vec function $w(\mathbf{r})$. At the maximum it is the spin Kramer function $K[\beta, \boldsymbol{\alpha}; v]$. Second part of the theorem is formulated for the spin-canonical ensemble: at a given inverse temperature $\beta$ and a 2vec electron density $\rho(\mathbf{r})$, the dimensionless potential $w_0(\mathbf{r})$ is determined uniquely. This potential $w_0$ minimizes w.r.t. $w$ the $\{X[\beta, w] + \int [w\rho]\}$ functional at fixed $\beta$ and fixed the 2vec electron density $\rho(\mathbf{r})$. At the minimum it is the spin Massieu universal functional $G[\beta, \rho]$. So, finally, the $w_0 \leftrightarrow \rho_0$ mapping is proven.*

Equivalently, in the energy representation (i.e., using Gibbs-Helmholtz fns.):

**Theorem 2.** *In the spin-grand-canonical ensemble at a given inverse temperature $\beta$ and in contact with a reservoir of the electrons characterized by a 2vec $\boldsymbol{\mu}$, the 2vec equilibrium electron density $\rho_0(\mathbf{r})$ is determined uniquely by the 2vec modified potential $u(\mathbf{r}) = \boldsymbol{\mu} - v(\mathbf{r})$. This $\rho_0$ minimizes w.r.t. $\rho$ the $\{F[\beta, \rho] - \int [u\rho]\}$ functional at fixed $\beta$ and fixed the 2vec function $u(\mathbf{r})$. At the minimum it is the spin grand potential $\Omega[\beta, \boldsymbol{\mu}; v]$. Second part of the theorem is formulated for the spin-canonical ensemble: at a given inverse temperature $\beta$ and a*

*2vec electron density* $\rho(\mathbf{r})$, *the potential* $\mathbf{u}_0(\mathbf{r})$ *is determined uniquely. This potential* $\mathbf{u}_0$ *maximizes w.r.t.* $\mathbf{u}$ *the* $\{B[\beta,\mathbf{u}]+\int[\mathbf{u}\rho]\}$ *functional at fixed* $\beta$ *and fixed the 2vec electron density* $\rho(\mathbf{r})$. *At the maximum it is the spin canonical universal functional* $F[\beta,\rho]$. *So, finally, the* $\mathbf{u}_0 \leftrightarrow \rho_0$ *mapping is proven.*

We would like to close this Section with some general remarks:

(i) The equivalence of the above two theorems (at finite temperature) follows from the equivalence, demonstrated in Paper I, of the $(M,m)$ Massieu function and the $(M,m)$ Gibbs-Helmholtz function as equilibrium-state descriptors of a given thermodynamic system.

(ii) For the $\{\hat{F}_{int},\hat{\rho}(\mathbf{r})\}$ system investigated alone, the one-to-one functional dependence between the 2vec function $\rho(\mathbf{r})$ (at equilibrium) and $\mathbf{w}(\mathbf{r})$ is established at fixed $\beta$. When the equilibrium state of this system is required to be identical with the equilibrium state of the $\{\hat{H},\hat{\mathcal{N}}\}$ system, the 2vec "local source" $\mathbf{w}(\mathbf{r})$ becomes a sum of the 2vec potential $\mathbf{v}(\mathbf{r})$ (present in $\hat{H}$) multiplied by $\beta$, and the 2vec "global source" $\boldsymbol{\alpha}$, as shown in Eq.(72). Now, this sum is in one-to-one relation with the equilibrium density $\rho(\mathbf{r})$. These relations hold in the case of a finite as well as vanishing magnetic component $B_z(\mathbf{r})$ of $\mathbf{v}(\mathbf{r})$.

(iii) As well known, measurable results are independent of an arbitrary constant shift applied to the scalar potential used in calculations. Is this true for the presented here theory of the equilibrium state? Consider the fully-open state of the $\{\hat{H}[\mathbf{v}],\hat{\mathcal{N}}\}$ system characterized at the equilibrium by the sources $\{\beta,\boldsymbol{\alpha}\}$. To see the role of the shift, let us replace the 2vec potential $\mathbf{v}(\mathbf{r})$ in $\hat{H}[\mathbf{v}]$ by the shifted (by a real 2vec constant $\mathbf{c}=(c_N,c_S)$) potential $\mathbf{v}'(\mathbf{r})=\mathbf{v}(\mathbf{r})+\mathbf{c}$. To retain the same equilibrium density matrix operator, the modifications of the potential $\mathbf{v}\to\mathbf{v}'$ should be accompanied by such modification of the sources $\{\beta,\alpha\} \rightarrow \{\beta',\alpha'\}$ that $\hat{\tilde{\varGamma}}_{eq}[\beta,\alpha;v] = \hat{\tilde{\varGamma}}_{eq}[\beta',\alpha';v']$. According to Eqs.(5), (4) this means that the conditions operators must be the same

$$-\beta\hat{H}[v] - \alpha\hat{\mathcal{N}} = -\beta'\hat{H}[v'] - \alpha'\hat{\mathcal{N}}. \tag{91}$$

Since $\hat{H}[v] = \hat{F}_{int} + \int [v\hat{\rho}]$ and $\int [v'\hat{\rho}] = \int [v\hat{\rho}] + c\hat{\mathcal{N}}$, we find from Eq.(91)

$$\beta = \beta', \qquad \alpha' = \alpha - \beta c. \tag{92}$$

It is easy to verify that the local source $w(\mathbf{r})$ calculated in Eq.(72) from $\{\beta,\alpha,v\}$ remains the same when calculated from $\{\beta',\alpha',v'\}$. Therefore the measurable object – the equilibrium density – $\rho(\mathbf{r}) = \tilde{\rho}[\beta,w](\mathbf{r})$ remains the same.

## VII. CONCLUSIONS

The generalization of DFT to (diagonal) spin DFT (SDFT) at finite temperature was proposed. Based on the properties of the basic Massieu function/functional, its Legendere transforms (various Massieu functions/functionals) and corresponding Massieu-Planck transforms (various Gibbs-Helmholtz functions/functionals), the equilibrium characteristics of two thermodynamic systems (introduced in Sec.II) were discussed. The equilibrium states under the spin-grand-canonical condition and the spin-canonical condition were considered in detail. State functions in both ensembles were precisely defined, their properties and the properties of their first and second derivatives were analyzed. The conditions for the equivalence between two systems for various state fns./fnls (i.e. Legendre and Massieu-Planck transforms) were derived and thoroughly discussed in the case of the spin-grand-canonical ensemble and the spin-canonical ensemble. Finally, based on these equivalence conditions the extension of the Hohenberg-Kohn theorem to the thermodynamic spin-density functional theory was performed both in the entropy (Massieu fns.) and energy (Gibbs-Helmholtz fns.) representations. Obtained results provide a rigorous mathematical foundation for obtaining the spin-density functional theory at zero temperature limit and for determining the chemical reactivity indices in the thermodynamic spin-density functional theory. This will be the subject of the next paper in the series.[60]

## ACKNOWLEDGMENT

The authors acknowledge support of Ministry of Science and Higher Education under Grant No. 1 T09A 025 30 and the International PhD Projects Programme of the Foundation for Polish Science, cofinanced from European Regional Development Fund within Innovative Economy Operational Programme "Grants for Innovation". Special thanks are due to Professor Paul W. Ayers for his valuable comments and beneficial remarks on the manuscript.

## APPENDIX. LIEB-VALONE FUNCTIONAL AND ITS EXTENSION.

The traditional zero-temperature DFT can be conveniently formulated in terms of the Lieb-Valone universal internal energy fnl. (see, e.g., Eq.(2.29) of Ref.[57] or Eq.(2.106) of Ref.[58])

$$F_L[\rho] = \inf \left\{ \langle \hat{F}_{int} \rangle_{\hat{\Gamma}} \, \middle| \, \hat{\Gamma}, \langle \hat{\rho}(\mathbf{r}) \rangle_{\hat{\Gamma}} = \rho(\mathbf{r}) \right\}. \quad (A1)$$

Here assumed that $\int [\rho] = \mathcal{N}$ is an integer, while the DM operator

$$\hat{\Gamma} = \left\{ \sum_K |\Psi_K\rangle g_K \langle \Psi_K| \, \middle| \, \langle \Psi_K|\Psi_L\rangle = \delta_{KL}, g_K \geq 0, \sum_K g_K = 1 \right\}. \quad (A2)$$

acts in the $\mathcal{N}$-electron Hilbert space, i.e., consisting of state functions $|\Psi_K\rangle$ satisfying

$$\hat{\mathcal{N}}|\Psi_K\rangle = \mathcal{N}|\Psi_K\rangle, \quad (A3)$$

with the same integer $\mathcal{N}$.

For the application to the zero-temperature (diagonal) SDFT, corresponding to the SDFT considered in the present paper at finite $\beta$, the natural extension of $F_{\mathrm{L}}[\rho]$ defined by Eq.(A1) is

$$F_{\mathrm{L}}[\boldsymbol{\rho}] = \inf\left\{\langle\hat{F}_{\mathrm{int}}\rangle_{\hat{\Gamma}} \big| \hat{\Gamma}, \langle\hat{\boldsymbol{\rho}}(\mathbf{r})\rangle_{\hat{\Gamma}} = \boldsymbol{\rho}(\mathbf{r})\right\}, \qquad (A4)$$

where the argument of the fnl. $F_{\mathrm{L}}$ is now the 2vec $\boldsymbol{\rho}(\mathbf{r}) = (\rho_{\mathrm{N}}(\mathbf{r}), \rho_{\mathrm{S}}(\mathbf{r}))$ assumed to satisfy $\int[\rho_{\mathrm{N}}] = \mathcal{N}$ and $\int[\rho_{\mathrm{S}}] = \mathcal{S}$; the pair of integers $(\mathcal{N}, \mathcal{S})$ must be admissible (see Sec.II), while the DM operator $\hat{\Gamma}$ occurring in Eq.(A4) acts in a subspace (sector) of the $\mathcal{N}$-electron Hilbert space consisting of state function $|\Psi_K\rangle$ satisfying

$$(\hat{\mathcal{N}}, \hat{\mathcal{S}})|\Psi_K\rangle = (\mathcal{N}, \mathcal{S})|\Psi_K\rangle, \qquad (A5)$$

with the same $(\mathcal{N}, \mathcal{S})$.

**Table I**. Legendre transforms, their conditions operators and corresponding Massieu-Plack transforms for the $\{\hat{H},\hat{\mathcal{N}}\}$ system (dependence on $\hat{H},\hat{\mathcal{N}}$ is suppressed for brevity and replaced by the dependence on $v$, the arguments of an object are often indicated as the superscript). Note the modified order of operators $\{\hat{H},\hat{\mathcal{S}},\hat{\mathcal{N}}\} \Rightarrow (j)$ in 3rd row. Note relation $\alpha = -\beta\mu$.

| Legendre transform (Massieu fn.) | conditions operator | Massieu-Planck transform (Gibbs-Helmholtz fn.) |
|---|---|---|
| the spin Kramer function $K$ <br><br> $\breve{\Theta}^{3,3}[\beta,\alpha;v] \equiv K[\beta,\alpha;v]$ <br><br> $= S^{\text{BGS}}\left[\hat{\tilde{\Gamma}}_{\text{eq}}^{[\beta,\alpha;v]}\right] - \beta \breve{E}^{[\beta,\alpha;v]} - \alpha\breve{\mathcal{N}}^{[\beta,\alpha;v]}$ <br><br> $= Y\left[\beta,\breve{\mathcal{N}}^{[\beta,\alpha;v]};v\right] - \alpha\breve{\mathcal{N}}^{[\beta,\alpha;v]}$ | $\hat{\tilde{O}}^{3,3}[\beta,\alpha;v] = -\beta\hat{H}[v] - \alpha\hat{\mathcal{N}}$ | the spin grand potential $\Omega$ <br><br> $\breve{\Upsilon}^{3,3}[\beta,\mu;v] \equiv \Omega[\beta,\mu;v]$ <br><br> $= -\beta^{-1}S^{\text{BGS}}\left[\hat{\tilde{\Gamma}}_{\text{eq}}^{[\beta,\mu;v]}\right] + \breve{E}^{[\beta,\mu;v]} - \mu\breve{\mathcal{N}}^{[\beta,\mu;v]}$ <br><br> $= A\left[\beta,\breve{\mathcal{N}}^{[\beta,\mu;v]};v\right] - \mu\breve{\mathcal{N}}^{[\beta,\mu;v]}$ |
| $\breve{\Theta}^{3,2}[\beta,\alpha_{\text{N}},\mathcal{S};v] = K\left[\beta,\alpha_{\text{N}},\breve{\alpha}_{\text{S}}^{[\beta,\alpha_{\text{N}},\mathcal{S};v]};v\right] + \mathcal{S}\breve{\alpha}_{\text{S}}^{[\beta,\alpha_{\text{N}},\mathcal{S};v]}$ <br><br> $= S^{\text{BGS}}\left[\hat{\tilde{\Gamma}}_{\text{eq}}^{[\beta,\alpha_{\text{N}},\mathcal{S};v]}\right] - \beta\breve{E}^{[\beta,\alpha_{\text{N}},\mathcal{S};v]} - \alpha_{\text{N}}\breve{\mathcal{N}}^{[\beta,\alpha_{\text{N}},\mathcal{S};v]}$ <br><br> $= Y\left[\beta,\breve{\mathcal{N}}^{[\beta,\alpha_{\text{N}},\mathcal{S};v]},\mathcal{S};v\right] - \alpha_{\text{N}}\breve{\mathcal{N}}^{[\beta,\alpha_{\text{N}},\mathcal{S};v]}$ | $\hat{\tilde{O}}^{3,2}[\beta,\alpha_{\text{N}},\mathcal{S};v] =$ <br><br> $-\beta\hat{H}[v] - \alpha_{\text{N}}\hat{\mathcal{N}} - \breve{\alpha}_{\text{S}}^{[\beta,\alpha_{\text{N}},\mathcal{S};v]}(\hat{\mathcal{S}} - \mathcal{S})$ | $\breve{\Upsilon}^{3,2}[\beta,\mu_{\text{N}},\mathcal{S};v] = \Omega\left[\beta,\mu_{\text{N}},\breve{\mu}_{\text{S}}^{[\beta,\mu_{\text{N}},\mathcal{S};v]};v\right] + \mathcal{S}\breve{\mu}_{\text{S}}^{[\beta,\mu_{\text{N}},\mathcal{S};v]}$ <br><br> $= -\beta^{-1}S^{\text{BGS}}\left[\hat{\tilde{\Gamma}}_{\text{eq}}^{[\beta,\mu_{\text{N}},\mathcal{S};v]}\right] + \breve{E}^{[\beta,\mu_{\text{N}},\mathcal{S};v]} - \mu_{\text{N}}\breve{\mathcal{N}}^{[\beta,\mu_{\text{N}},\mathcal{S};v]}$ <br><br> $= A\left[\beta,\breve{\mathcal{N}}^{[\beta,\mu_{\text{N}},\mathcal{S};v]},\mathcal{S};v\right] - \mu_{\text{N}}\breve{\mathcal{N}}^{[\beta,\mu_{\text{N}},\mathcal{S};v]}$ |
| $\breve{\Theta}^{3,2}_{(j)}[\beta,\alpha_{\text{S}},\mathcal{N};v] = K\left[\beta,\breve{\alpha}_{\text{N}}^{[\beta,\alpha_{\text{S}},\mathcal{N};v]},\alpha_{\text{S}};v\right] + \mathcal{N}\breve{\alpha}_{\text{N}}^{[\beta,\alpha_{\text{S}},\mathcal{N};v]}$ <br><br> $= S^{\text{BGS}}\left[\hat{\tilde{\Gamma}}_{\text{eq}}^{[\beta,\alpha_{\text{S}},\mathcal{N};v]}\right] - \beta\breve{E}^{[\beta,\alpha_{\text{S}},\mathcal{N};v]} - \alpha_{\text{S}}\breve{\mathcal{S}}^{[\beta,\alpha_{\text{S}},\mathcal{N};v]}$ <br><br> $= Y\left[\beta,\mathcal{N},\breve{S}^{[\beta,\alpha_{\text{S}},\mathcal{N};v]}\right] - \alpha_{\text{S}}\breve{\mathcal{S}}^{[\beta,\alpha_{\text{S}},\mathcal{N};v]}$ | $\hat{\tilde{O}}^{3,2}_{(j)}[\beta,\alpha_{\text{S}},\mathcal{N};v] =$ <br><br> $-\beta\hat{H}[v] - \alpha_{\text{S}}\hat{\mathcal{S}} - \breve{\alpha}_{\text{N}}^{[\beta,\alpha_{\text{S}},\mathcal{N};v]}(\hat{\mathcal{N}} - \mathcal{N})$ | $\breve{\Upsilon}^{3,2}_{(j)}[\beta,\mu_{\text{S}},\mathcal{N};v] = \Omega\left[\beta,\breve{\mu}_{\text{N}}^{[\beta,\mu_{\text{S}},\mathcal{N};v]},\mu_{\text{S}};v\right] + \mathcal{N}\breve{\mu}_{\text{N}}^{[\beta,\mu_{\text{S}},\mathcal{N};v]}$ <br><br> $= -\beta^{-1}S^{\text{BGS}}\left[\hat{\tilde{\Gamma}}_{\text{eq}}^{[\beta,\mu_{\text{S}},\mathcal{N};v]}\right] + \breve{E}^{[\beta,\mu_{\text{S}},\mathcal{N};v]} - \mu_{\text{S}}\breve{\mathcal{S}}^{[\beta,\mu_{\text{S}},\mathcal{N};v]}$ <br><br> $= A\left[\beta,\mathcal{N},\breve{S}^{[\beta,\mu_{\text{S}},\mathcal{N};v]};v\right] - \mu_{\text{S}}\breve{\mathcal{S}}^{[\beta,\mu_{\text{S}},\mathcal{N};v]}$ |
| the spin Massieu function $Y$ <br><br> $\breve{\Theta}^{3,1}[\beta,\mathcal{N};v] \equiv Y[\beta,\mathcal{N};v]$ <br><br> $= K\left[\beta,\breve{\alpha}^{[\beta,\mathcal{N};v]};v\right] + \mathcal{N}\breve{\alpha}^{[\beta,\mathcal{N};v]}$ <br><br> $= S^{\text{BGS}}\left[\hat{\tilde{\Gamma}}_{\text{eq}}^{[\beta,\mathcal{N};v]}\right] - \beta\breve{E}^{[\beta,\mathcal{N};v]} = -\beta A[[\beta,\mathcal{N};v]]$ | $\hat{\tilde{O}}^{3,1}[\beta,\mathcal{N};v] =$ <br><br> $-\beta\hat{H}[v] - \breve{\alpha}^{[\beta,\mathcal{N};v]}(\hat{\mathcal{N}} - \mathcal{N})$ | the spin Helmholtz function $A$ <br><br> $\breve{\Upsilon}^{3,1}[\beta,\mathcal{N};v] \equiv A[\beta,\mathcal{N};v]$ <br><br> $= \Omega\left[\beta,\breve{\mu}^{[\beta,\mathcal{N};v]};v\right] + \mathcal{N}\breve{\mu}^{[\beta,\mathcal{N};v]}$ <br><br> $= -\beta^{-1}S^{\text{BGS}}\left[\hat{\tilde{\Gamma}}_{\text{eq}}^{[\beta,\mathcal{N};v]}\right] + \breve{E}^{[\beta,\mathcal{N};v]} = -\beta^{-1}Y[\beta,\mathcal{N};v]$ |

| the entropy $S^{\text{BGS}}$ $$\breve{\Theta}^{3,0}[E,\mathcal{N};v] = S^{\text{BGS}}\left[\hat{\tilde{\Gamma}}_{\text{eq}}^{[E,\mathcal{N};v]}\right]$$ $$= K\left[\breve{\beta}^{[E,\mathcal{N};v]},\breve{\alpha}^{[E,\mathcal{N};v]};v\right] + E\breve{\beta}^{[E,\mathcal{N};v]} + \mathcal{N}\breve{\alpha}^{[E,\mathcal{N};v]}$$ | $$\hat{\tilde{O}}^{3,0}[E,\mathcal{N};v] = -\breve{\beta}^{[E,\mathcal{N};v]}\left(\hat{H}[v]-E\right)$$ $$-\breve{\alpha}^{[E,\mathcal{N};v]}\left(\hat{\mathcal{N}}-\mathcal{N}\right)$$ | |
|---|---|---|

**Table II**. Legendre transforms, their conditions operators and corresponding Massieu-Plack transforms for the $\{\hat{F}_{int},\hat{\rho}\}$ system (dependence on $\hat{F}_{int},\hat{\rho}$ is suppressed for brevity, the arguments of an object are often indicated as the superscript, $\beta$ means $b$). Note the modified order of operators $\{\hat{F}_{int},\tilde{\rho}_S,\tilde{\rho}_N\} \Rightarrow (j)$ in 3rd row. Nore realtion $w(\mathbf{r}) = -\beta u$, and the notation $\int[f] \equiv \int f(\mathbf{r})d\mathbf{r}$.

| Legendre transform (Massieu functional) | conditions operator | Massieu-Planck transform (Gibbs-Helmholtz functional) |
|---|---|---|
| the spin Kramer functional $X$ <br> $\tilde{\Theta}^{3,3}[\beta,w] \equiv X[\beta,w] = G\left[\beta,\tilde{\rho}^{[\beta,w]}\right] - \int\left[w\tilde{\rho}^{[\beta,w]}\right]$ <br> $= S^{BGS}\left[\hat{\tilde{\Gamma}}^{[\beta,w]}_{eq}\right] - \beta\tilde{F}^{[\beta,w]}_{int} - \int\left[w\tilde{\rho}^{[\beta,w]}\right]$ | $\hat{\tilde{O}}^{3,3}[\beta,w] = -\beta\hat{F}_{int} - \int[w\hat{\rho}]$ | the spin grand functional $B$ <br> $\tilde{\Upsilon}^{3,3}[\beta,u] \equiv B[\beta,u] = F\left[\beta,\tilde{\rho}^{[\beta,u]}\right] - \int\left[u\tilde{\rho}^{[\beta,u]}\right]$ <br> $= -\beta^{-1}S^{BGS}\left[\hat{\tilde{\Gamma}}^{[\beta,u]}_{eq}\right] + \tilde{F}^{[\beta,u]}_{int} - \int\left[u\tilde{\rho}^{[\beta,u]}\right]$ |
| $\tilde{\Theta}^{3,2}[\beta,w_N,\rho_S] = G\left[\beta,\tilde{\rho}^{[\beta,w_N,\rho_S]}_N,\rho_S\right] - \int\left[w_N\tilde{\rho}^{[\beta,w_N,\rho_S]}_N\right]$ <br> $= S^{BGS}\left[\hat{\tilde{\Gamma}}^{[\beta,w_N,\rho_S]}_{eq}\right] - \beta\tilde{F}^{[\beta,w_N,\rho_S]}_{int} - \int\left[w_N\tilde{\rho}^{[\beta,w_N,\rho_S]}_N\right]$ | $\hat{\tilde{O}}^{3,2}[\beta,w_N,\rho_S] = -\beta\hat{F}_{int} - \int[w_N\hat{\rho}_N]$ <br> $-\int\left[\tilde{w}^{[\beta,w_N,\rho_S]}_S(\hat{\rho}_S - \rho_S)\right]$ | $\tilde{\Upsilon}^{3,2}[\beta,u_N,\rho_S] = F\left[\beta,\tilde{\rho}^{[\tilde{\beta},u_N,\rho_S]}_N,\rho_S\right] - \int\left[u_N\tilde{\rho}^{[\beta,u_N,\rho_S]}_N\right]$ <br> $= -\beta^{-1}S^{BGS}\left[\hat{\tilde{\Gamma}}^{[\beta,u_N,\rho_S]}_{eq}\right] + \tilde{F}^{[\beta,u_N,\rho_S]}_{int} - \int\left[u_N\tilde{\rho}^{[\beta,u_N,\rho_S]}_N\right]$ |
| $\tilde{\Theta}^{3,2}_{(j)}[\beta,w_S,\rho_N] = G\left[\beta,\rho_N,\tilde{\rho}^{[\beta,w_S,\rho_N]}_S\right] - \int\left[w_S\tilde{\rho}^{[\beta,w_S,\rho_N]}_S\right]$ <br> $= S^{BGS}\left[\hat{\tilde{\Gamma}}^{[\beta,w_S,\rho_N]}_{eq}\right] - \beta\tilde{F}^{[\beta,w_S,\rho_N]}_{int} - \int\left[w_S\tilde{\rho}^{[\beta,w_S,\rho_N]}_S\right]$ | $\hat{\tilde{O}}^{3,2}_{(j)}[\beta,w_S,\rho_N] = -\tilde{\beta}\hat{F} - \int[w_S\hat{\rho}_S]$ <br> $-\int\left[\tilde{w}^{[\beta,w_S,\rho_N]}_N(\hat{\rho}_N - \rho_N)\right]$ | $\tilde{\Upsilon}^{3,2}_{(j)}[\beta,u_S,\rho_N] = F\left[\beta,\rho_N,\rho^{[\beta,u_S,\rho_N]}_S\right] - \int\left[u_S\tilde{\rho}^{[\beta,u_S,\rho_N]}_S\right]$ <br> $= -\beta^{-1}S^{BGS}\left[\hat{\tilde{\Gamma}}^{[\beta,u_S,\rho_N]}_{eq}\right] + \tilde{F}^{[\beta,u_S,\rho_N]}_{int} - \int\left[u_S\tilde{\rho}^{[\beta,u_S,\rho_N]}_S\right]$ |
| the spin Massieu universal functional $G$ <br> $\tilde{\Theta}^{3,1}[\beta,\rho] \equiv G[\beta,\rho] = S^{BGS}\left[\hat{\tilde{\Gamma}}^{[\beta,\rho]}_{eq}\right] - \beta\tilde{F}^{[\beta,\rho]}_{int}$ <br> $= -\beta F[\beta,\rho]$ | $\hat{\tilde{O}}^{3,1}[\beta,\rho] = -\beta\hat{F}_{int} - \int\left[\tilde{w}^{[\beta,\rho]}(\hat{\rho} - \rho)\right]$ | the spin canonical universal functional $F$ <br> $\tilde{\Upsilon}^{3,1}[\beta,\rho] \equiv F[\beta,\rho] = -\beta^{-1}S^{BGS}\left[\hat{\tilde{\Gamma}}^{[\beta,\rho]}_{eq}\right] + \tilde{F}^{[\beta,\rho]}_{int}$ <br> $= -\beta^{-1}G[\beta,\rho]$ |
| the entropy $S^{BGS}$ <br> $\tilde{\Theta}^{3,0}[F_{int},\rho] \equiv S^{BGS}\left[\hat{\tilde{\Gamma}}^{[F_{int},\rho]}_{eq}\right]$ <br> $= G\left[\tilde{\beta}^{[F_{int},\rho]},\rho\right] + \tilde{\beta}^{[F_{int},\rho]}F_{int}$ | $\hat{\tilde{O}}^{3,0}[F_{int},\rho] = -\tilde{\beta}^{[F_{int},\rho]}\left(\hat{F}_{int} - F_{int}\right)$ <br> $-\int\left[\tilde{w}^{[F_{int},\rho]}(\hat{\rho} - \rho)\right]$ | ✕ |

**Table III**. The equivalence conditions between two systems for the Legendre and the Massieu-Planck transforms for different ensembles. Note relations $\alpha = -\beta\mu$, $w(\mathbf{r}) = -\beta u(\mathbf{r})$.

| Legendre transform | Massieu-Planck transform |
|---|---|
| the spin grand canonical ensemble | |
| $\beta = b$ | |
| $\beta v(\mathbf{r}) + \alpha = w(\mathbf{r})$ | $v(\mathbf{r}) - \mu = -u(\mathbf{r})$ |
| the spin canonical ensemble at given spin number | |
| $\beta = b$ | |
| $\mathcal{S} = \int \rho_S(\mathbf{r}) d\mathbf{r}$ | |
| $\begin{cases} \beta v_{\text{ext}}(\mathbf{r}) + \alpha_N = w_N(\mathbf{r}) \\ \beta B_z(\mathbf{r}) + \breve{\alpha}_S^{[\beta,\alpha_N,\mathcal{S};v]} = \tilde{w}_S^{[\beta,w_N,\rho_S]}(\mathbf{r}) \end{cases}$ | $v_{\text{ext}}(\mathbf{r}) - \mu_N = -u_N(\mathbf{r})$ <br> $B_z(\mathbf{r}) - \breve{\mu}_S^{[\beta,\mu_N,\mathcal{S};v]} = -\tilde{u}_S^{[\beta,u_N,\rho_S]}(\mathbf{r})$ |
| the spin canonical ensemble at given particle number | |
| $\beta = b$ | |
| $\mathcal{N} = \int \rho_N(\mathbf{r}) d\mathbf{r}$ | |
| $\beta v_{\text{ext}}(\mathbf{r}) + \breve{\alpha}_N^{[\beta,\alpha_S,\mathcal{N};v]} = \tilde{w}_N^{[\beta,w_S,\rho_N]}(\mathbf{r})$ <br> $\beta B_z(\mathbf{r}) + \alpha_S = w_S(\mathbf{r})$ | $v_{\text{ext}}(\mathbf{r}) - \breve{\mu}_N^{[\beta,\mu_S,\mathcal{N};v]} = -\tilde{u}_N^{[\beta,u_S,\rho_N]}(\mathbf{r})$ <br> $B_z(\mathbf{r}) - \mu_S = -u_S(\mathbf{r})$ |
| the spin canonical ensemble at given both particle and spin numbers | |
| $\beta = b$ | |
| $\mathcal{N} = \int \rho(\mathbf{r}) d\mathbf{r}$ | |
| $\beta v(\mathbf{r}) + \breve{\alpha}^{[\beta,\mathcal{N};v]} = \tilde{w}^{[\beta,\rho]}(\mathbf{r})$ | $v(\mathbf{r}) - \breve{\mu}^{[\beta,\mathcal{N};v]} = -\tilde{u}^{[\beta,\rho]}(\mathbf{r})$ |
| the spin microcanonical ensemble | |
| $\breve{\beta}^{[E,\mathcal{N};v]} = \tilde{\beta}^{[F_{\text{int}},\rho]}$ <br> $\mathcal{N} = \int [\rho]$ <br> $E = F_{\text{int}} + \int [v\rho]$ <br> $\breve{\beta}^{[E,\mathcal{N};v]} v(\mathbf{r}) + \breve{\alpha}^{[E,\mathcal{N};v]} = \tilde{w}^{[F_{\text{int}},\rho]}(\mathbf{r})$ | |

**Figure 1**. All possible independent Legendre transforms $\breve{\Theta}^{(M,m)}$ for the $\{\hat{H},\hat{\mathcal{N}},\hat{\mathcal{S}}\}$ system. Besides the initial ordered set of the system operators, two reordered sets are introduced: $\{\hat{\mathcal{N}},\hat{\mathcal{S}},\hat{H}\} \Rightarrow (l)$, $\{\hat{\mathcal{S}},\hat{H},\hat{\mathcal{N}}\} \Rightarrow (k)$.

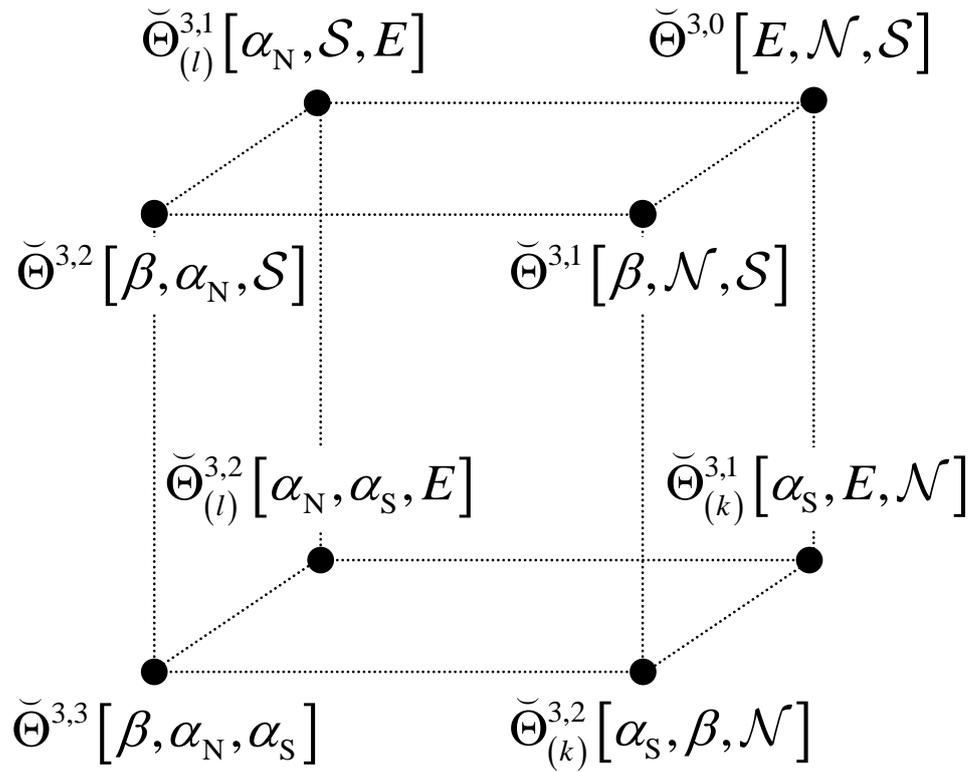